\documentclass{iopjournal}
\pdfoutput=1
\usepackage[T1]{fontenc}
\usepackage[utf8]{inputenc}
\usepackage{lmodern}
\usepackage{amsmath,amssymb,mathrsfs,bm}
\usepackage{booktabs,array,tabularx,longtable}
\usepackage{microtype}
\usepackage{etoolbox}
\usepackage[caption=false,font=footnotesize,labelfont=bf]{subfig}
\patchcmd{\articletype}{Journal Name}{Classical and Quantum Gravity}{}{\PackageError{cqg}{Journal placeholder was not found}{Use the supplied iopjournal class.}}
\patchcmd{\articletype}{dd Month yyyy}{\textemdash}{}{}
\patchcmd{\articletype}{dd Month yyyy}{\textemdash}{}{}
\usepackage[numbers,sort&compress]{natbib}
\usepackage{url}

\hypersetup{pdftitle={Extended Hamiltonian thermodynamics and Carroll contractions of AdS black holes},pdfauthor={Yingnan Xu and Shuangshuang Chu},colorlinks=true,allcolors=blue}
\numberwithin{equation}{section}
\newcommand{\ren}{\mathrm{ren}}
\newcommand{\GC}{G_C}
\newcommand{\eps}{\varepsilon}
\newcommand{\dd}{\mathrm{d}}
\newcommand{\Lie}{\mathcal L}
\newcommand{\vol}{\boldsymbol\epsilon}
\newcommand{\Thgrav}{\boldsymbol\Theta}
\newcommand{\HH}{\mathcal H}
\newcommand{\ii}{\iota}
\newcommand{\Order}{\mathcal O}
\newcommand{\Tr}{\operatorname{tr}}
\begin{document}
\articletype{Paper}

\title{Extended Hamiltonian thermodynamics and Carroll contractions of AdS black holes}

\author{Yingnan Xu$^{1,2,*}$\orcid{0009-0005-8296-0850} and Shuangshuang Chu$^3$\orcid{0009-0004-4396-582X}}

\noindent\affil{$^1$Zhongtai Securities Institute for Financial Studies, Shandong University, Jinan, Shandong 250014, China}

\noindent\affil{$^2$Department of Physics, Southern Methodist University, Dallas, Texas 75206, USA}

\noindent\affil{$^3$School of Statistics, Dongbei University of Finance and Economics, Dalian, Liaoning 116025, China}

\noindent\affil{$^*$Author to whom any correspondence should be addressed.}

\email{yingnanx@mail.smu.edu; shuangshchu@gmail.com}

\noindent\keywords{Carrollian gravity, black-hole thermodynamics, cosmological constant, Hamiltonian constraints, covariant phase space}

\begin{abstract}
We study extended anti-de Sitter black-hole thermodynamics under correlated contractions of the time generator and Newton coupling. We relate the contraction parameter to physical units and distinguish variations within an equilibrium black-hole family from changes along the contraction. Starting from the magnetic Carroll Hamiltonian action, we derive the boundary identity with a variable cosmological constant and evaluate its energy, horizon and pressure contributions for Schwarzschild--anti-de Sitter geometries. We verify the spatial and lapse equations, retaining the boost connection in the coframe formulation. A three-form potential supplies a canonical variable conjugate to the cosmological constant. We compare this static boundary mechanics with Lorentzian thermodynamic contractions, whose common scaling of the work terms selects a finite Carroll branch. On this branch, temperature vanishes and entropy diverges while their conjugate product remains finite. The pressure and volume scalings depend on the chosen physical normalization. Charged, rotating and higher-dimensional families provide further thermodynamic tests, including the asymptotic frame and full thermodynamic volume for rotation. We use symbolic identities and numerical differentiation to verify the thermodynamic variations, resolve the conjugate contributions and check their scaling behaviour. These constructions relate extended static Carroll boundary mechanics to Lorentzian thermodynamic contractions and identify the phase-space conditions under which they correspond.
\end{abstract}

% 01_introduction.tex
\section{Introduction}
\label{sec:introduction}

In this paper, we relate the thermodynamic variation of an anti-de Sitter black hole to the Hamiltonian boundary variation of magnetic Carroll gravity with a variable cosmological constant. The relation depends on the normalization of time translations, the gravitational coupling and the boundary conditions used to define energy. A contraction of the Lorentzian thermodynamic identity determines the scaling of the work terms. Its Carroll gravitational interpretation also requires a limiting action, canonical data satisfying the constraints of that action, and the resulting boundary variation. We establish this relation for the static Schwarzschild--anti-de Sitter family and use charged and rotating Lorentzian families to examine how additional thermodynamic variables enter the correspondence.

The Carroll geometry used here originates from contractions of the Poincar\'e group \cite{LevyLeblond:1965,SenGupta:1966,Duval_2014}. Its formulations through gauging, non-Lorentzian differential geometry and expansions of general relativity supply the geometric framework for our calculation \cite{Hartong:2015xda,Hansen:2021fxi,Bergshoeff:2022eog,FigueroaOFarrill:2022mcy,Hartong:2022lsy}. In magnetic Carroll gravity, the Hamiltonian retains the spatial curvature potential and imposes a constraint on the evolution of the spatial metric. We use the established Hamiltonian contraction and coframe action developed in \cite{HenneauxSalgadoRebolledo:2021,Campoleoni:2022wmf}. For negative cosmological constant, P\'erez constructed the asymptotic Hamiltonian charges and analysed the Schwarzschild--anti-de Sitter solution \cite{Perez:2022carrollads}. These results supply the gravitational theory and static boundary charge needed for the present calculation. We extend the corresponding static boundary variation to neighbouring values of the cosmological constant and follow it through the Lorentzian contraction.

The intrinsic Carroll black holes of Ecker et al.\ provide a direct comparison for this construction \cite{Ecker:2023uwm}. Their analysis identifies Carroll thermal geometries and Carroll extremal surfaces, includes Schwarzschild, Reissner--Nordstr\"om and lower-dimensional examples, and derives a thermodynamic law from boundary charges. Their work also relates Lorentzian and Carrollian Poisson-sigma models through a target-space diffeomorphism. Their treatment also explains the normalization required to assign thermodynamic dimensions to entropy. We adopt their fixed-radius magnetic scaling as the physical reference for our static representative. We evaluate the pressure conjugate with its subtraction at the anti-de Sitter boundary, and trace the map between canonical Carroll quantities and Lorentzian work terms as the Hamiltonian generator is rescaled. These are the additional elements of the comparison developed in this paper. Throughout this comparison, the entropy of each finite Lorentzian theory retains its own normalization, while the rescaled entropy coordinate enters the limiting thermodynamic identity.

In related work, Tadros and Kol\'a\v r studied Carrollian limits of quadratic gravity and Schwarzschild geometries with cosmological constant \cite{Tadros:2023,Tadros:2024}. Their analysis establishes the vanishing physical temperature, divergent entropy and thermodynamic response of the magnetic limit. We use these established thermodynamic results to interpret the magnetic limit and examine how the chosen normalization enters its pressure work. For this comparison, we relate the physical Newton constant, the coupling multiplying the dimensionless gravitational action, and the pressure entering the thermodynamic identity. These relations explain how divergent pressure in one convention can coexist with finite physical pressure in the magnetic scaling. Treating entropy and pressure variations together also allows us to specify which quantities remain fixed in each response function and to follow that choice consistently through the contraction.

Rotation brings an additional constraint into the comparison. Kol\'a\v r, Kubiz\v n\'ak and Tadros restrict stationary axisymmetric magnetic Carroll solutions in more than three spacetime dimensions to static configurations, subject to the stated topological qualification \cite{Kolar:2025rotating}. Their analysis of the Carroll BTZ black hole already includes an extended thermodynamic law with pressure and volume. We use the Kerr--anti-de Sitter family to examine the contraction of Lorentzian work terms involving angular momentum. The calculation distinguishes the angular velocity in Boyer--Lindquist coordinates from that measured in a frame that is nonrotating at infinity, and retains the rotation contribution to thermodynamic volume. We then identify the momentum scaling of a fixed-rotation family, which determines how this Lorentzian thermodynamic test is related to the canonical magnetic limit.

The pressure variation is formulated within extended black-hole mechanics, where mass is an enthalpy and the cosmological constant determines pressure \cite{Kastor:2009wy,Cvetic:2010jb,Dolan:2011xt,Kubiznak:2012wp,Kubiznak:2016qmn}. We use the Noether-charge construction of Wald and Iyer \cite{Wald:1993nt,Iyer:1994ys} and its extension to variable couplings \cite{Xiao:2023lap}. The cosmological constant can label neighbouring theories or occur as an integration constant in an enlarged theory with a differential-form potential. Both descriptions lead to a volume contribution, which we derive with the same boundary normalization. In the enlarged theory, the additional canonical pair survives the magnetic contraction and gives the variable cosmological constant a phase-space interpretation. The normalization of the time generator remains fixed during thermodynamic variations in either description, so the pressure variation has a prescribed Hamiltonian meaning.

These questions also arise in the wider study of gravitational boundaries. Carroll structures organize data at null infinity and at black-hole horizons \cite{Ciambelli:2019lap,Donnay:2019jiz,Herfray:2021qgr}. We also draw on the horizon analysis of \cite{Donnay:2019jiz}, where the null Raychaudhuri and Damour equations are expressed as Carrollian conservation laws. In that setting, horizon symmetries include supertranslations and superrotations of Bondi--Metzner--Sachs (BMS) type with conserved horizon charges. The relations among horizon dynamics, Carrollian fluids and boundary symplectic structures provide further motivation for examining the conjugate boundary variables \cite{Freidel:2022vjq,Freidel:2022bai,RedondoYuste:2022czg,Husnugil:2025carroll,Bagchi:2025vri,Ruzziconi:2026review}. In holography, related structures appear in flat-space gravity and celestial amplitudes \cite{Ciambelli:2018wre,Pasterski:2021rjz,Raclariu:2021zjz,Pasterski:2021fjn,Donnay:2022aba,Donnay:2022wvx,Mason:2023mti,Liu:2024carrollamp,Nguyen:2025lectures}. We regard Carrollian conformal theories and theories on null manifolds as possible settings for a future statistical interpretation of these variables \cite{Bagchi:2019clu,Bagchi:2019xfx,Bagchi:2024xfg,Aggarwal:2025universal}. The role of the present classical calculation is to determine the normalization and conjugate quantities that such a statistical interpretation would need to reproduce.

The choice of physical quantities held fixed is also central when comparing different gravitational limits. We keep Planck's constant fixed here, while tantum gravity involves a joint scaling of physical constants \cite{Ecker:2025tantum}. The gravitational thermodynamic variables must likewise be related to boundary pressure and central-charge variations through a specified holographic ensemble \cite{Karch_2015,Mancilla:2024spp,Mann:2025chemistry,Borsboom:2026ash}. Nonrelativistic brane limits and Newton--Cartan holography offer further examples of the role played by the variables retained in a contraction \cite{Guijosa:2023qym,Lambert:2024m2,Lambert:2024d3,PhysRevLett.133.151601,Fontanella:2024rvb,Guijosa:2025nrh,fontanella2026triality,Fontanella:2026gaq}. We retain these broader connections as motivation and carry out the derivation in the bulk gravitational theory, where the coupling, time generator and boundary variations can be specified together.

The paper is organized around this sequence of constructions. Section~\ref{sec:setup} defines the geometric family, physical units and permitted variations, and Section~\ref{sec:iw} derives the extended Lorentzian boundary identity. In Section~\ref{sec:contractions}, we classify the thermodynamic contractions and introduce finite conjugate variables. Section~\ref{sec:magnetic} develops their gravitational interpretation through the magnetic Hamiltonian action, the static field equations and the extended boundary terms. We then analyse Schwarzschild thermodynamics and its limiting sectors in Section~\ref{sec:schwarzschild}. Charge, rotation and spacetime dimension are treated in Sections~\ref{sec:charged}, \ref{sec:rotating} and \ref{sec:dimensions}, followed by the numerical checks in Section~\ref{sec:numerics}. Section~\ref{sec:conclusion} discusses the results and their scope. The appendices complete the derivation with the coframe symplectic structure, the differential-form realization of a variable cosmological constant, and the treatment of changes in clock normalization.

% 02_setup.tex
\section{Geometric family, physical units and variations}
\label{sec:setup}

\subsection{The contraction and the thermodynamic variation}

Our starting point is the four-dimensional Schwarzschild--anti-de Sitter family. We abbreviate anti-de Sitter space as AdS and write its line element as
\begin{equation}
 \dd s^2=-c^2 f(r)\dd t^2+\frac{\dd r^2}{f(r)}+r^2\dd\Omega_2^2,
 \qquad f(r)=1-\frac{r_0}{r}+\frac{r^2}{\ell^2},
 \qquad r_0=r_h\left(1+\frac{r_h^2}{\ell^2}\right).
 \label{eq:sads-family}
\end{equation}
Here $c>0$ is a dimensionless contraction parameter, $\ell>0$ is the AdS radius, and $r_h>0$ is the zero of $f$. We work in the exterior, taking the bifurcation surface of each finite-$c$ solution as the inner boundary. The spacetime orientation agrees with $\dd t\wedge\dd r\wedge\dd\theta\wedge\dd\phi$. Throughout the neutral calculation, $r_0$ denotes the coefficient of the inverse-radius term. Its relation to the horizon radius $r_h$ is fixed by the displayed horizon equation, so the geometric meanings of $r_0$ and $r_h$ remain distinct.

We define the finite Lorentzian clock, the chosen time generator and the coupling by
\begin{equation}
 \tau=ct,\qquad
 \xi_\lambda=c^{-\alpha}\partial_t=c^{1-\alpha}\partial_\tau,
 \qquad G(c)=c^\gamma\GC,\qquad \lambda(c)=c^{-\alpha}.
 \label{eq:scaling-definition}
\end{equation}
The constant $\GC>0$ and the real exponents $\alpha$ and $\gamma$ are held fixed. At each $c$, we define charges relative to a specified asymptotic AdS time translation and assign zero charge to the corresponding AdS reference. The notation $H_\lambda$ denotes the energy conjugate to $\xi_\lambda$; we retain $H_t$ and $H_\tau$ for the charges associated with $\partial_t$ and $\partial_\tau$, respectively. These subscripts record the time normalization used to measure energy.

We separate motion along the contraction curve from thermodynamic variations at fixed contraction data. Along the curve in $c$, we hold $(r_h,\ell)$ at finite nonzero values, thereby fixing $r_0$ through Eq.~\eqref{eq:sads-family}. At each finite $c$, we allow independent variations $\delta r_h$ and $\delta\ell$ in the extended equilibrium family. The corresponding restrictions and induced variation of the radius coefficient are
\begin{equation}
 \delta c=\delta\alpha=\delta\gamma=\delta\GC=0,
 \qquad
 \delta r_0=\left(1+\frac{3r_h^2}{\ell^2}\right)\delta r_h
             -\frac{2r_h^3}{\ell^3}\delta\ell.
 \label{eq:allowed-variation}
\end{equation}
Thus $\delta G=0$ within a thermodynamic variation, even when $G(c)$ tends to zero along the contraction curve. The geometric charge and rotation parameters vary within their respective families under the same convention. We hold them fixed along $c$ unless an additional scaling is stated. Our calculations use smooth families with finite limiting coefficients. Special states at which an individual coefficient vanishes are treated separately from the scaling of the complete thermodynamic differential in a neighbourhood of that state.

For functions on an equilibrium family, we write this variation as the
tangent operator
\begin{equation}
 \delta=\delta r_h\,\partial_{r_h}
       +\delta\ell\,\partial_\ell
       +\delta q\,\partial_q
       +\delta a\,\partial_a .
 \label{eq:thermodynamic-tangent-operator}
\end{equation}
We include the charge and rotation directions only in families containing
those parameters. With the contraction exponents already fixed,
$\delta c=\delta\GC=0$ keeps the Newton coupling and clock normalization
fixed during differentiation, while $\delta\ell$ induces a variation of
the cosmological constant. In the Einstein formulation used in
Section~\ref{sec:iw}, this variation connects equilibrium solutions of
theories with neighbouring values of $\Lambda$. Appendix~\ref{app:threeform}
realizes the same extended variation within one enlarged theory:
a three-form potential makes $\Lambda$ an integration constant whose
value can differ between solutions. We use the tangent operator above
for the geometric variables in both descriptions.

In discussing the original Carroll limit, we regard Eq.~\eqref{eq:sads-family} as a family of tensors expressed in a fixed $t$ chart. Its temporal component tends to zero in that chart for every value of $\alpha$, which labels the generator normalization. To compare the clocks associated with different generators, we introduce $s=c^\alpha t$ and $\eps=c^{1-\alpha}$. Then $\xi_\lambda=\partial_s$, $\tau=\eps s$, and the temporal part of the metric becomes $-\eps^2 f\dd s^2$. The Carroll branch in this adapted clock is defined by $\alpha<1$, for which $\eps\to0$. At $\alpha=1$, the adapted clock coincides with $\tau$ and the metric retains its ordinary Lorentzian form. This convention specifies both the coordinate basis in which the metric limit is taken and the normalization of the time translation used in the thermodynamic calculation.

The norm of the selected generator relates these two descriptions
directly:
\begin{equation}
 g(c)(\xi_\lambda,\xi_\lambda)
 =c^{-2\alpha}g_{tt}(c)
 =-c^{2(1-\alpha)}f(r)
 =-\eps^2 f(r).
 \label{eq:selected-generator-norm}
\end{equation}
Outside the horizon, the generator is timelike at every finite $c$.
Its norm tends to zero for $\alpha<1$, remains $-f(r)$ for
$\alpha=1$, and grows in magnitude for $\alpha>1$.
Since $\xi_\lambda=\partial_s$, the same norm is the temporal metric
coefficient in the adapted clock; the fixed-$t$ metric component
retains its separate factor $c^2$.

\subsection{Relation to the physical Newton constant}

To connect the coupling $G$ to physical quantities, we state its dimensional convention in four dimensions. Let $G_N$ denote the physical Newton constant, $c_{\rm phys}$ the physical speed of light, and $\hbar$ the fixed Planck constant. We express the gravitational action in units of $\hbar$ and use a temporal coordinate with dimensions of length. The coupling in that dimensionless action also determines the dimensionless entropy:
\begin{equation}
 G=\frac{\hbar G_N}{c_{\rm phys}^{3}},
 \qquad
 \frac{I_{\rm EH}}{\hbar}
 =\frac{1}{16\pi G}\int (R-2\Lambda)\vol,
 \qquad
 S=\frac{S_{\rm phys}}{k_B}=\frac{A}{4G}.
 \label{eq:physical-coupling}
\end{equation}
The coupling $G$ therefore has dimensions of area and equals the squared Planck length. In the subsequent action and charge calculations, we use $\hbar=k_B=1$. Keeping these factors in the present subsection allows us to track how the thermodynamic quantities in that convention are converted into physical observables.

For a physical interpretation of $c$, we choose a fixed reference speed $c_*$ and set $c_{\rm phys}=cc_*$. If $t_{\rm phys}$ is the physical time, we take $t=c_*t_{\rm phys}$ and $\tau=c_{\rm phys}t_{\rm phys}$. For a Schwarzschild coefficient $r_0$, we then have
\begin{equation}
 r_0=\frac{2G_NM_{\rm phys}}{c_{\rm phys}^{2}},
 \qquad
 E_{\rm phys}=M_{\rm phys}c_{\rm phys}^{2}
             =\frac{c_{\rm phys}^{4}r_0}{2G_N}.
 \label{eq:physical-radius-energy}
\end{equation}
Keeping $r_0$ fixed along the contraction correlates the physical mass and Newton constant. In the magnetic convention of \cite{Ecker:2023uwm,Tadros:2024}, we keep $G_M=G_Nc_{\rm phys}^{-4}$ fixed. We obtain
\begin{equation}
 G_N=c_{\rm phys}^{4}G_M,
 \qquad G=c\,\hbar c_*G_M,
 \qquad \GC=\hbar c_*G_M,
 \qquad M_{\rm phys}=\frac{r_0}{2G_Mc_{\rm phys}^{2}}.
 \label{eq:magnetic-unit-map}
\end{equation}
The fixed-generator representative $(\alpha,\gamma)=(0,1)$ consequently reproduces the usual physical magnetic scaling. Its physical energy remains $r_0/(2G_M)$, while its mass grows as $c_{\rm phys}^{-2}$. For a general curve $G=c^\gamma\GC$, the conversion instead gives $G_N=c_*^3\GC c^{\gamma+3}/\hbar$. Holding $G_M$ fixed selects $\gamma=1$ under the stated identification of $c$ with physical speed. Other exponents specify different paths for the physical constants, together with their chosen coupling and generator normalizations. We use the magnetic representative as the reference when interpreting the general thermodynamic classification.

Along the fixed-radius contraction curve, we can derive the mass scaling
directly by logarithmically differentiating
Eq.~\eqref{eq:physical-radius-energy}. With $c_{\rm phys}=cc_*$ and
$c_*$ fixed, we obtain
\begin{equation}
 \frac{\dd\ln G_N}{\dd\ln c}
 +\frac{\dd\ln M_{\rm phys}}{\dd\ln c}-2=0.
 \label{eq:fixed-radius-logarithmic-scaling}
\end{equation}
These derivatives follow the contraction curve at fixed $r_0$.
The coupling map gives $\dd\ln G_N/\dd\ln c=\gamma+3$ and hence
$\dd\ln M_{\rm phys}/\dd\ln c=-\gamma-1$.
For the magnetic representative, $\gamma=1$ gives the mass exponent
$-2$ in Eq.~\eqref{eq:magnetic-unit-map}.
The thermodynamic variation $\delta$ continues to act at fixed
contraction data as specified in Eq.~\eqref{eq:allowed-variation}.

Like energy, pressure must be converted from the normalization used in the dimensionless action to physical units. For the static family, the complete set of conversion rules is
\begin{equation}
 \begin{aligned}
 E_{\rm phys}&=\hbar c_*H_t,&
 k_BT_{\rm phys}&=\hbar c_*T_t,\\
 P_{\rm phys}&=\hbar c_{\rm phys}P
       =-\frac{\Lambda c_{\rm phys}^{4}}{8\pi G_N},&
 V_{\rm phys}&=c^{-1}V_t.
 \end{aligned}
 \label{eq:physical-thermodynamic-map}
\end{equation}
At fixed $c$ and fixed physical constants, these rules convert the generator-normalized work term into $V_{\rm phys}\delta P_{\rm phys}$. In the magnetic representative, $P$ diverges and $V_t$ vanishes, while $P_{\rm phys}$ and $V_{\rm phys}$ remain finite. These differing limits follow from converting both conjugate variables using Eq.~\eqref{eq:physical-thermodynamic-map}. In physical units, the temperature vanishes and the entropy in units of $k_B$ diverges. Section~\ref{sec:contractions} introduces finite conjugate coordinates for the boundary identity while retaining this physical interpretation of the individual Lorentzian quantities.

Throughout this map, $\hbar$ remains fixed. The tantum-gravity scaling holds $\hbar c_{\rm phys}$ fixed together with $G_M$ \cite{Ecker:2025tantum}, thereby changing the entropy normalization in Eq.~\eqref{eq:physical-coupling}. We use the fixed-$\hbar$ convention to compare the finite Lorentzian theories considered in this paper. For the extension to other spacetime dimensions, we state the dimension of the gravitational coupling separately. The four-dimensional mass conversion derived here supplies the physical reference for the neutral magnetic representative.

% 03_covariant.tex
\section{The extended Lorentzian boundary identity}
\label{sec:iw}

\subsection{Cosmological-constant source and thermodynamic volume}

The action in Eq.~\eqref{eq:physical-coupling} defines the Einstein Lagrangian form used throughout this section. At fixed $G$, varying the fields and the cosmological constant gives
\begin{equation}
 \delta\mathbf L
 =\mathbf E^{\mu\nu}\delta g_{\mu\nu}
   +\dd\Thgrav(g,\delta g)
   -\frac{\delta\Lambda}{8\pi G}\vol.
 \label{eq:lagrangian-variation}
\end{equation}
We require the background metric to satisfy the Einstein equation and the perturbation to satisfy the corresponding linearized equation, including the source proportional to $\delta\Lambda$. For a Killing vector $\xi$ held fixed during this variation, the surface form is $\mathbf k_\xi=\delta\mathbf Q_\xi-\ii_\xi\Thgrav$, where $\mathbf Q_\xi$ denotes the Noether charge. The differential and integrated versions of the Iyer--Wald identity are then \cite{Wald:1993nt,Iyer:1994ys,Kastor:2009wy,Xiao:2023lap}
\begin{equation}
 \dd\mathbf k_\xi=\frac{\delta\Lambda}{8\pi G}\ii_\xi\vol,
 \qquad
 \int_{S_R}\mathbf k_\xi-\int_B\mathbf k_\xi
 =\frac{\delta\Lambda}{8\pi G}\int_{\Sigma_R}\ii_\xi\vol.
 \label{eq:iw-source-identity}
\end{equation}
Here $B$ is the bifurcation surface and $S_R$ is an outer cutoff surface. In the displayed difference, the orientation of $B$ is induced by increasing radius. We retain the cutoff while specifying the surface and bulk subtractions, so that both the energy variation and the cosmological-constant source are evaluated with the same asymptotic prescription before the limit is taken.

For a static solution, we choose $\xi$ to generate the horizon and normalize it relative to a fixed asymptotic time. We obtain $\int_B\mathbf k_\xi=\kappa_\xi\delta A/(8\pi G)=T_\xi\delta S$, with $T_\xi=\kappa_\xi/(2\pi)$. We define
\begin{equation}
 P=-\frac{\Lambda}{8\pi G},\qquad
 \delta P=-\frac{\delta\Lambda}{8\pi G},\qquad
 V_\xi=-\int_\Sigma^{\ren}\ii_\xi\vol.
 \label{eq:pressure-volume-definition}
\end{equation}

In four dimensions, $\Lambda=-3/\ell^2$ and the fixed-coupling
thermodynamic variation give
\begin{equation}
 \delta P=-\frac{2P}{\ell}\,\delta\ell
          =-\frac{3}{4\pi c^\gamma\GC\ell^3}\,\delta\ell .
 \label{eq:pressure-component-of-variation}
\end{equation}
The factor $c^{-\gamma}$ sets the normalization of the pressure
differential generated by the tangent operator in
Eq.~\eqref{eq:thermodynamic-tangent-operator}.
The geometric dependence of this differential is determined by the
variation of the AdS radius.

After subtracting the AdS reference in Eq.~\eqref{eq:iw-source-identity}, we identify the outer surface variation with $\delta H_\xi$ and obtain
\begin{equation}
 \delta H_\xi=T_\xi\delta S+V_\xi\delta P.
 \label{eq:static-extended-law}
\end{equation}
In the extended ensemble, $H_\xi$ has the interpretation of enthalpy \cite{Kastor:2009wy,Dolan:2011xt,Kubiznak:2016qmn}. The volume in Eq.~\eqref{eq:pressure-volume-definition} is a subtracted integral weighted by the lapse. The proper spatial volume of the exterior instead uses the spatial measure without that weight and diverges at infinity. Our pressure coefficient is determined by the subtracted lapse-weighted integral entering the Hamiltonian identity.

For the static subtraction, we use the same asymptotic clock on the black-hole and AdS families. A constant generator factor $b$ gives the contraction $\ii_{b\partial_\tau}\vol=b r^2\dd r\wedge\dd\Omega_2$ of the volume form in Eq.~\eqref{eq:sads-family}, expressed in the $\tau$ coordinate. The reference and black-hole integrals consequently differ only in their radial domains:
\begin{equation}
 \int_\Sigma^{\ren}\ii_{b\partial_\tau}\vol
 =\lim_{R\to\infty}\frac{4\pi b}{3}
       \big[(R^3-r_h^3)-R^3\big]
 =-\frac{4\pi b r_h^3}{3}.
 \label{eq:static-volume-subtraction}
\end{equation}
The energy variation is subtracted using this same clock and reference. As $\ell$ varies, we compare the black hole with the neighbouring AdS geometry at the corresponding value of $\ell$. This procedure defines an extended family of asymptotic boundary conditions. We fix the finite normalization of its charges by assigning zero energy to AdS and the usual static AdS mass to the black hole.

Matching proper time at a finite cutoff provides a useful comparison with this prescription. Such matching multiplies the AdS lapse by $b_R=\sqrt{f(R)/f_0(R)}$, where $f_0=1+R^2/\ell^2$. Its expansion is $b_R=1-r_0\ell^2/(2R^3)+\Order(R^{-5})$. Multiplying the reference bulk volume by $b_R$ leaves a finite contribution $2\pi b r_0\ell^2/3$ in addition to Eq.~\eqref{eq:static-volume-subtraction}. The outer surface variation must then be transformed consistently with the lapse. We use a common asymptotic clock throughout, which assigns the bulk integral and energy variation a single reference prescription. For a rotating solution, the asymptotic frame contributes further terms to this comparison; Section~\ref{sec:rotating} includes them in the calculation of thermodynamic volume.

\subsection{Generator dependence and additional work terms}

For a constant $b$ with $\delta b=0$, the Noether charge and surface form satisfy $\mathbf Q_{b\xi}=b\mathbf Q_\xi$ and $\mathbf k_{b\xi}=b\mathbf k_\xi$. With a reference prescription transformed by the same factor, we have
\begin{equation}
 \delta H_{b\xi}=b\delta H_\xi,\qquad
 T_{b\xi}=bT_\xi,\qquad V_{b\xi}=bV_\xi.
 \label{eq:generator-linearity}
\end{equation}
Entropy and pressure remain unchanged under this rescaling of the generator. They are determined by the area and cosmological constant at a given coupling, independently of the time normalization. Combining Eq.~\eqref{eq:generator-linearity} with the change of $G$ between finite-$c$ theories then gives the contraction weights of all terms in the thermodynamic variation.

For a stationary solution, we distinguish the asymptotic time translation $k$ from the horizon generator $\chi=k+\Omega\psi$, where $\psi$ generates rotations and its Hamiltonian charge is $-J$. The infinity term evaluated with the reference-solution generator $\chi$ is the one-form $\delta H_k-\Omega\delta J$. We include the electromagnetic potential difference with a compatible gauge choice when charge is present. The resulting thermodynamic law is
\begin{equation}
 \delta H_k=T_\chi\delta S+\Omega\delta J+\Phi\delta Q+V_k\delta P.
 \label{eq:stationary-extended-law}
\end{equation}
The subscript $k$ identifies the asymptotic time normalization used for energy and its conjugates, with $H_k$ reserved for the charge of that time translation. We hold the generator at its reference-solution value when evaluating the variation. The thermodynamic coefficients can vary between neighbouring solutions while the boundary prescription remains fixed.

When a coordinate representation makes $\xi$ depend on the solution parameters, we use the surface form
\begin{equation}
 \mathbf k_\xi=\delta\mathbf Q_\xi-\mathbf Q_{\delta\xi}-\ii_\xi\Thgrav.
 \label{eq:adjusted-surface-form}
\end{equation}
This surface form evaluates the field variation at a prescribed generator. In a rotating AdS chart, the boundary conditions must also correspond to the nonrotating frame used to define the mass; we implement both requirements in Section~\ref{sec:rotating}. The contraction factor $\lambda(c)$ remains fixed during thermodynamic variations. Appendix~\ref{app:clock} retains the algebraic contribution from changing this normalization and relates it to a variation of the external clock data.

% 04_contractions.tex
\section{Thermodynamic contractions and finite conjugate variables}
\label{sec:contractions}

\subsection{Scaling of the thermodynamic differential}

We now apply Eq.~\eqref{eq:generator-linearity} to geometric families and tangent vectors that are independent of $c$. A hat denotes a thermodynamic quantity evaluated with coupling $\GC$ and the finite-clock normalization. Thus $\widehat S=A/(4\GC)$ and $\widehat P=-\Lambda/(8\pi\GC)$. The generator factor is $\eps=c^{1-\alpha}$, as defined in Section~\ref{sec:setup}, while $\eta=c^{1-\alpha-\gamma}$ denotes the common weight of the work terms. Table~\ref{tab:weights} collects the resulting map. Its electric-charge entry uses an Einstein--Maxwell normalization with a common inverse gravitational coupling, which we specify in Section~\ref{sec:charged} before applying the map to charged solutions.

\begin{table}[tb]
\caption{Contraction weights for fixed geometric parameters. Hatted quantities are evaluated at $G=\GC$ with the finite-clock normalization. The charge and rotation entries apply when their work terms are present. All factors are constant during a thermodynamic variation.}
\label{tab:weights}
\centering
\renewcommand{\arraystretch}{1.3}
\begin{tabular}{lll}
\toprule
Quantity & Relation to the hatted variable & Origin of the factor\\
\midrule
$H_\lambda$ & $\eta\widehat H$ & Generator and coupling\\
$T_\lambda,V_\lambda,\Phi_\lambda,\Omega_\lambda$
 & $\eps(\widehat T,\widehat V,\widehat\Phi,\widehat\Omega)$ & Generator\\
$S,P,Q,J$ & $c^{-\gamma}(\widehat S,\widehat P,\widehat Q,\widehat J)$ & Coupling\\
\bottomrule
\end{tabular}
\end{table}

We obtain an equality between thermodynamic one-forms on the geometric parameter space,
\begin{equation}
 \begin{aligned}
 \delta H_\lambda
 &=\eta\,\delta\widehat H,\\
 T_\lambda\delta S+V_\lambda\delta P
       +\Phi_\lambda\delta Q+\Omega_\lambda\delta J
 &=\eta\left(\widehat T\delta\widehat S
       +\widehat V\delta\widehat P
       +\widehat\Phi\delta\widehat Q
       +\widehat\Omega\delta\widehat J\right).
 \end{aligned}
 \label{eq:work-form-scaling}
\end{equation}
The hatted quantities satisfy the same thermodynamic identity at $G=\GC$. Generator linearity and the coupling weight together determine the exponent in Eq.~\eqref{eq:work-form-scaling}. We use this exponent to select the normalization that retains a finite thermodynamic differential. We then use the action calculation in Section~\ref{sec:magnetic} to identify the limiting gravitational theory and the canonical conditions defining the selected sector.

Requiring a finite nonzero limit of the differential on the full geometric family selects the branch
\begin{equation}
 \alpha+\gamma=1,
 \qquad \alpha<1\ \Longrightarrow\ \gamma>0,
 \qquad G=\eps\GC.
 \label{eq:finite-branch}
\end{equation}
The restriction $\alpha<1$ applies when we use the adapted Carroll clock. For $\alpha+\gamma<1$, the differential tends to zero; for $\alpha+\gamma>1$, its nonzero coefficients grow in magnitude. Individual terms may vanish at extremality or at zero charge or rotation, and particular tangent vectors may annihilate the energy differential. We therefore classify the one-form on a neighbourhood of the state, including independent variations of its geometric parameters. This formulation retains the distinction between a property of the full family and a vanishing coefficient at a special state.

Table~\ref{tab:sectors} collects the distinguished limits. On the finite Carroll branch, $T_\lambda$ and $V_\lambda$ tend to zero, while $S$ and $P$ grow as $\eps^{-1}$. Their products in Eq.~\eqref{eq:work-form-scaling} remain finite. The metric limit in the original $t$ chart follows from Eq.~\eqref{eq:sads-family}. The table describes the limiting metric in the adapted $s$ chart. Specifying both charts allows us to track the geometric limit and the thermodynamic generator normalization as separate operations.

\begin{table}[tb]
\caption{Distinguished contraction sectors. The entries for the work terms refer to nonzero coefficients on the geometric parameter space. We describe the metric in the clock $s=c^\alpha t$ associated with the chosen generator.}
\label{tab:sectors}
\centering
\renewcommand{\arraystretch}{1.3}
\begin{tabularx}{\linewidth}{@{}>{\raggedright\arraybackslash}p{0.23\linewidth}>{\raggedright\arraybackslash}p{0.22\linewidth}>{\raggedright\arraybackslash}X@{}}
\toprule
Sector & Exponents & Limiting behaviour\\
\midrule
Fixed coupling and generator & $(0,0)$ & Carroll metric; $H_t,T_t,V_t\to0$; entropy and pressure fixed.\\
Finite Carroll branch & $\alpha<1$, $\gamma=1-\alpha$ & Carroll metric; finite work terms; $T_\lambda,V_\lambda\to0$ and $S,P\to\infty$.\\
Physical magnetic representative & $(0,1)$ & Fixed $G_M$ in Eq.~\eqref{eq:magnetic-unit-map}; finite physical energy and pressure.\\
Lorentzian endpoint & $(1,0)$ & $s=\tau$; ordinary Lorentzian metric and finite-clock thermodynamics.\\
Growing generator clock & $\alpha>1$ & The temporal coefficient in the adapted clock grows; the work terms retain their separate $\eta$ scaling.\\
\bottomrule
\end{tabularx}
\end{table}

\subsection{Finite variables, integrability and physical interpretation}

On the finite branch defined by Eq.~\eqref{eq:finite-branch}, the energy satisfies $H_\lambda=\widehat H$. Multiplying $S$ and $P$ by $\eps$ gives finite entropy and pressure coordinates, while dividing $T_\lambda$ and $V_\lambda$ by the same factor gives their finite conjugates, as recorded in Table~\ref{tab:weights}. This change of variables preserves each conjugate product:
\begin{equation}
 T_\lambda\delta S=\widehat T\delta\widehat S,
 \qquad V_\lambda\delta P=\widehat V\delta\widehat P,
 \qquad
 \Phi_\lambda\delta Q+\Omega_\lambda\delta J
 =\widehat\Phi\delta\widehat Q+\widehat\Omega\delta\widehat J.
 \label{eq:finite-pairings}
\end{equation}
The hatted entropy is a normalized thermodynamic coordinate defined with coupling $\GC$. At every finite $c$, the unrescaled entropy remains $A/(4G)$, and its divergence is part of the physical fixed-$\hbar$ limit. Similarly, $\widehat T$ is the finite coefficient conjugate to $\widehat S$, with the physical temperature obtained through Eq.~\eqref{eq:physical-thermodynamic-map}. These definitions connect the boundary normalization of Carroll entropy discussed in \cite{Ecker:2023uwm} to the physical temperature and entropy behaviour analysed in \cite{Tadros:2024}, while keeping the meaning of each entropy variable fixed throughout the calculation.

The same change of variables determines the conjugate area elements in thermodynamic variable space. Since the contraction factors are fixed under $\delta$, the finite branch preserves $\delta T_\lambda\wedge\delta S=\delta\widehat T\wedge\delta\widehat S$ and the corresponding pressure, electric and rotational pairings. These identities describe a transformation of thermodynamic state variables. Below, we derive the gravitational presymplectic form from the action on the space of constrained gravitational fields. The distinction between these spaces specifies the role of each construction when relating the thermodynamic contraction to magnetic gravity.

An integrable mass on the original Lorentzian family gives an integrable hatted energy. At fixed $c$, multiplication of the complete one-form by $\eta$ commutes with the exterior derivative on geometric parameter space. The AdS reference fixes the integration constant of the resulting potential, whose derivatives we verify for each family. Along the finite Carroll branch, changing $\alpha$ changes the rate at which the generator clock approaches its limit. For the same geometric data and $\GC$, the hatted thermodynamic functions coincide across these choices of clock. The different contraction rates therefore preserve the same normalized thermodynamic potential and its derivatives.

Under the physical-speed identification of Section~\ref{sec:setup}, the finite branch gives $E_{\rm phys}=\hbar c_*c^\alpha\widehat H$ and $P_{\rm phys}=\hbar c_*c^\alpha\widehat P$. Physical energy and pressure therefore have finite nonzero limits on the fixed-$G_M$ representative $\alpha=0$. For $0<\alpha<1$, both quantities tend to zero, while for $\alpha<0$ their nonzero values grow in magnitude. Other points of the finite branch preserve the energy conjugate to their own chosen generator. These conversion factors determine how the formal thermodynamic contraction is interpreted relative to a fixed physical clock.

At fixed AdS scale, the inverse area coupling can also be related to the number of degrees of freedom in a proposed holographic description \cite{Karch_2015,Mancilla:2024spp,Borsboom:2026ash}. Such an interpretation requires an identified boundary theory and thermodynamic ensemble. The present bulk calculation determines the conjugate variables and their contraction weights, thereby specifying the gravitational quantities to be matched by that boundary description.

We formulate this interpretation in four bulk dimensions by considering
an identified boundary theory with
$N_{\rm eff}\propto\ell^2/G$ at fixed AdS scale and fixed ensemble.
The coupling map then gives $N_{\rm eff}\propto c^{-\gamma}$.
Under this identification, the finite Carroll branch with
$\gamma>0$ combines the gravitational contraction with a
large-$N_{\rm eff}$ limit.
The proportionality and the ensemble are inputs from the boundary theory;
the bulk scaling determines the resulting dependence on the contraction
parameter.

% 05_magnetic.tex
\section{Magnetic Carroll action and static boundary mechanics}
\label{sec:magnetic}

\subsection{Canonical limit and momentum sector}

On the finite thermodynamic branch, we relate the normalized boundary variables to the magnetic Hamiltonian formulation of gravity and identify the canonical conditions supporting that relation \cite{HenneauxSalgadoRebolledo:2021,Campoleoni:2022wmf}. We use the generator coordinate $s$ and contraction parameter $\eps$ defined in Section~\ref{sec:setup}. In the ADM decomposition, $h_{ij}$ denotes the spatial metric, $N$ the lapse, $v^i$ the shift, and $\pi^{ij}$ the momentum density conjugate to $h_{ij}$. Taking the Lorentzian lapse to be $\eps N$ gives the metric
\begin{equation}
 \dd s_g^2=-\eps^2N^2\dd s^2
 +h_{ij}(\dd x^i+v^i\dd s)(\dd x^j+v^j\dd s).
 \label{eq:magnetic-adm-metric}
\end{equation}
On the branch in Eq.~\eqref{eq:finite-branch}, the finite-$c$ canonical action and its constraints are
\begin{equation}
 \begin{aligned}
 I_\eps&=\int\dd s\left[
 \int_\Sigma\dd^{D-1}x
 \left(\pi^{ij}\partial_s h_{ij}-N\HH_\eps-v^i\HH_i\right)
 -Q_{\partial\Sigma}\right],\\
 \HH_\eps&=
 \frac{16\pi\GC\eps^2}{\sqrt h}
 \left(\pi^{ij}\pi_{ij}-\frac{\pi^2}{D-2}\right)
 -\frac{\sqrt h}{16\pi\GC}(R[h]-2\Lambda),
 \qquad \HH_i=-2D_j\pi^j{}_i.
 \end{aligned}
 \label{eq:magnetic-parent-action}
\end{equation}
Here $\pi=h_{ij}\pi^{ij}$ and $D_i$ denotes the spatial Levi-Civita derivative. The total Hamiltonian consists of the bulk constraints and the surface term $Q_{\partial\Sigma}$, so the latter enters the action with a minus sign. Below, we determine its variation from the boundary contribution generated by varying the bulk Hamiltonian.

The powers of the contraction parameter in the parent action connect the thermodynamic scaling condition to the relative contributions of curvature and canonical momentum. Before the finite-branch condition is imposed, the curvature potential carries $c^{1-\alpha-\gamma}$ and the quadratic momentum term carries $c^{1-\alpha+\gamma}$. A finite nonzero curvature coefficient therefore requires the same condition that preserves the thermodynamic differential. On this line, the relative kinetic coefficient is $c^{2(1-\alpha)}=\eps^2$. At the finite-clock endpoint, both ADM contributions remain in the Hamiltonian. Along the adapted Carroll branch, the magnetic potential determines the limit provided the canonical momentum satisfies the regularity condition given below.

For this canonical contraction, we consider smooth families in which $h_{ij}$, $\pi^{ij}$, their variations and the spatial derivatives entering the action all remain finite. In terms of $\mathcal K_{ij}=(\partial_sh_{ij}-\Lie_vh_{ij})/(2N)$, the relation between momentum and metric evolution expresses the required scaling as
\begin{equation}
 \pi^{ij}=\frac{\sqrt h}{16\pi\GC\eps^2}
 \left(\mathcal K^{ij}-h^{ij}\mathcal K\right),
 \qquad \mathcal K_{ij}=\Order(\eps^2).
 \label{eq:magnetic-momentum-scaling}
\end{equation}
Setting $\eps=0$ in Eq.~\eqref{eq:magnetic-parent-action} gives the magnetic action $I_C$, whose lapse constraint we denote by $\HH_C$. The quadratic momentum term drops out, while momentum remains an independent canonical field. Its variation requires the spatial metric to evolve by the shift, thereby enforcing the magnetic evolution equation. The stated regularity condition therefore controls the limit of the action and canonical variables as well as that of the spatial geometry. The static family considered here has $\pi^{ij}=0$ at every finite $c$, so its contraction satisfies this condition throughout the approach to the limit.

At fixed $\GC$, we retain the bulk canonical potential and two-form,
\begin{equation}
 \Theta_{C,\Sigma}(\delta)=\int_\Sigma\dd^{D-1}x\,
       \pi^{ij}\delta h_{ij},
 \qquad
 \Omega_{C,\Sigma}(\delta_1,\delta_2)=
 \int_\Sigma\dd^{D-1}x\,
 (\delta_1\pi^{ij}\delta_2h_{ij}-\delta_2\pi^{ij}\delta_1h_{ij}).
 \label{eq:magnetic-canonical-symplectic}
\end{equation}

On a cutoff slice $\Sigma_R$, we express the boundary completion of the variational principle through the canonical potential. We denote the corner potential supplied by this completion by $\Theta_{\partial\Sigma_R}$. For commuting variations, the total potential and the associated two-form satisfy
\begin{equation}
 \begin{aligned}
 \Theta_{C,\Sigma_R}^{\rm tot}(\delta)
 &=\Theta_{C,\Sigma_R}(\delta)
   +\Theta_{\partial\Sigma_R}(\delta),\\
 \Omega_{C,\Sigma_R}^{\rm tot}(\delta_1,\delta_2)
 &=\Omega_{C,\Sigma_R}(\delta_1,\delta_2)
   +\delta_1\Theta_{\partial\Sigma_R}(\delta_2)
   -\delta_2\Theta_{\partial\Sigma_R}(\delta_1).
 \end{aligned}
 \label{eq:magnetic-boundary-presymplectic-completion}
\end{equation}
Here the bulk terms are the restrictions of
Eq.~\eqref{eq:magnetic-canonical-symplectic} to $\Sigma_R$.
The boundary action and allowed variations determine the corner
potential, including the case in which it vanishes.
The asymptotic limit of this completion is taken with the chosen
falloffs and boundary prescription. The static calculation below
evaluates the energy and horizon surface one-forms under the
common-clock AdS subtraction.

For the local constraint algebra, we restrict the variations to compact support, so the integrations by parts produce no surface contributions. We evaluate the asymptotic surface variations with the prescribed AdS subtraction and the boundary conditions of the static charge calculation. The support restriction therefore determines the local canonical algebra, while the asymptotic prescription determines the surface charges. Appendix~\ref{app:coframe} derives the same bulk form with the boost connection retained in the coframe action.

\subsection{Constraint algebra and static field equations}

We define $\mathscr D[X]=\int_\Sigma\pi^{ij}\Lie_Xh_{ij}$ and $\mathscr H_\eps[N]=\int_\Sigma N\HH_\eps$ for compactly supported smearings. With $\{h_{ij}(x),\pi^{kl}(y)\}=\delta^k_{(i}\delta^l_{j)}\delta^{D-1}(x-y)$, we obtain
\begin{equation}
 \begin{aligned}
 \{\mathscr D[X],\mathscr D[Y]\}&=\mathscr D[[X,Y]],\\
 \{\mathscr D[X],\mathscr H_\eps[N]\}&=
                  \mathscr H_\eps[\Lie_XN],\\
 \{\mathscr H_\eps[N],\mathscr H_\eps[M]\}&=
 \eps^2\mathscr D\!\left[h^{ij}(N\partial_jM-M\partial_jN)\right].
 \end{aligned}
 \label{eq:magnetic-constraint-algebra}
\end{equation}
The spatial diffeomorphism relations persist in the limit, while the lapse constraints commute because $\HH_C$ depends only on $h_{ij}$. Eq.~\eqref{eq:magnetic-constraint-algebra} thus gives the magnetic Carroll contraction of the local ADM algebra. Asymptotic charges additionally require allowed falloffs and surface improvements. The magnetic AdS analysis of \cite{Perez:2022carrollads} supplies the relevant setting for those boundary choices, while the compact support of the smearings above isolates the local constraint relations.

We denote the bulk coefficient in the variation of $\mathscr H_C[N]$ by
\begin{equation}
 \mathcal E_N^{ij}=\frac{\sqrt h}{16\pi\GC}
 \left[
 N\left(R^{ij}-\frac12h^{ij}(R-2\Lambda)\right)
 -D^iD^jN+h^{ij}\Delta N\right].
 \label{eq:magnetic-metric-variation}
\end{equation}
The Hamilton equations give $\partial_sh_{ij}=\Lie_vh_{ij}$ and $\partial_s\pi^{ij}=\Lie_v\pi^{ij}-\mathcal E_N^{ij}$. On a static slice with vanishing momentum and shift, we must therefore impose
\begin{equation}
 R[h]=2\Lambda,
 \qquad
 D_iD_jN=N\left(R_{ij}-\frac{2\Lambda}{D-2}h_{ij}\right).
 \label{eq:magnetic-static-equations}
\end{equation}
Taking the trace gives $\Delta N=-2\Lambda N/(D-2)$. We impose the scalar constraint from lapse variation together with the lapse equation required for the canonical momentum to remain stationary. These conditions agree with the static magnetic equations obtained from the expansion of general relativity \cite{Hansen:2021fxi} and provide the field equations used to check the spherical geometry below.

The Hessian terms in Eq.~\eqref{eq:magnetic-metric-variation} arise from integrating the derivative terms in $\delta R[h]$ twice against the lapse. Their presence in the momentum equation continues to constrain the lapse when the Carroll extrinsic curvature vanishes. Magnetic gravity therefore constrains the spatial geometry through both its curvature and the lapse, while retaining momentum as independent canonical data. In applying these equations to Schwarzschild, we use the zero-momentum static sector inherited from the Lorentzian family and require the metric to be independent of the selected clock.

We check the spherical family in $D\geq4$ dimensions by setting $n=D-2$, $h_{rr}=f^{-1}$ and $h_{AB}=r^2\gamma_{AB}$, with $\gamma_{AB}$ the unit sphere metric. We take $N=\sqrt f$ and $f=1-\mu r^{1-n}+r^2/\ell^2$, so $\mu=r_0$ in Eq.~\eqref{eq:sads-family}. We obtain
\begin{equation}
 \begin{gathered}
 \begin{aligned}
 R_{rr}&=-\frac{nf'}{2rf},&
 R_{AB}&=\left[(n-1)(1-f)-\frac{rf'}2\right]\gamma_{AB},\\
 D_rD_rN&=\frac{f''}{2\sqrt f},&
 D_AD_BN&=\frac{r\sqrt f f'}2\gamma_{AB},
 \end{aligned}
 \\
 R[h]=\frac{n}{r^2}\big[(n-1)(1-f)-rf'\big]
      =-\frac{n(n+1)}{\ell^2},
 \qquad
 \Delta N=\frac{\sqrt f}{2}\left(f''+\frac{nf'}r\right)
          =\frac{n+1}{\ell^2}N.
 \end{gathered}
 \label{eq:magnetic-spherical-curvature}
\end{equation}
Substituting $\Lambda=-n(n+1)/(2\ell^2)$ verifies the scalar constraint and both tensor components of Eq.~\eqref{eq:magnetic-static-equations}. The mass-dependent terms cancel from the scalar-curvature and lapse traces, but remain in the spatial Ricci tensor and lapse Hessian. The resulting geometry is the static magnetic Schwarzschild--AdS solution appearing in the existing Carroll constructions \cite{Hansen:2021fxi,Ecker:2023uwm,Tadros:2024}. This component calculation verifies the bulk equations on the static family, providing the conditions under which the remaining Hamiltonian variation reduces to boundary and cosmological-constant contributions.

The same spatial slice identifies the inner surface geometrically. Proper distance satisfies $\partial_\rho r=\sqrt f$, so the derivative of the sphere area with respect to proper distance vanishes at $r=r_h$. Extending the spatial solution across that surface produces the throat described by the Carroll extremal-surface construction in \cite{Ecker:2023uwm}. Its area enters the boundary variation below, and the lapse gradient at the surface determines the temperature coefficient for the chosen clock normalization. The same inner surface thus supplies both the geometric area and the lapse gradient that enter the thermodynamic boundary identity.

\subsection{Surface variation and the cosmological-constant term}

We write $q_{ij}=\delta h_{ij}$, $q^{ij}=h^{ik}h^{jl}q_{kl}$ and $q=h^{ij}q_{ij}$. Thus the raised tensor $q^{ij}$ differs in sign from the variation of the inverse metric. At fixed reference lapse, the constrained Hamiltonian varies as
\begin{equation}
 \begin{aligned}
 \delta\mathscr H_C[N]
 &=\int_\Sigma\dd^{D-1}x\,\mathcal E_N^{ij}q_{ij}
 -\frac{1}{16\pi\GC}\int_{\partial\Sigma}
      \dd^{D-2}x\sqrt\sigma\,n_i\mathcal B_N^i
 +\frac{\delta\Lambda}{8\pi\GC}\int_\Sigma\dd^{D-1}x\,N\sqrt h,\\
 \mathcal B_N^i
 &=N(D_jq^{ij}-D^iq)-(D_jN)q^{ij}+(D^iN)q.
 \end{aligned}
 \label{eq:magnetic-boundary-variation}
\end{equation}
Here $n_i$ is the outward unit normal and $\sigma$ is the induced metric. We complete the Hamiltonian by adding a surface term whose variation cancels the metric contribution at the boundary. Throughout this linearized calculation, we evaluate the lapse on the reference solution and hold its profile fixed. Contributions proportional to a lapse variation multiply a constraint and vanish on the solution family, consistently with the use of a prescribed generator in the surface identity.

To state the differentiability condition, let $\delta_h$ vary the
spatial metric at fixed $N$, $\GC$, $\Lambda$ and boundary sources.
For a boundary ensemble in which the surface one-form is
integrable, the Hamiltonian improvement is determined by
\begin{equation}
 \begin{aligned}
 \delta_h Q_{\partial\Sigma}[N]
 &=\frac{1}{16\pi\GC}
   \int_{\partial\Sigma}\dd^{D-2}x\sqrt{\sigma}\,
        n_i\mathcal B_N^i[\delta h],\\
 \delta_h\!\left(\mathscr H_C[N]+Q_{\partial\Sigma}[N]\right)
 &=\int_\Sigma\dd^{D-1}x\,\mathcal E_N^{ij}q_{ij}.
 \end{aligned}
 \label{eq:magnetic-boundary-improvement}
\end{equation}
This cancellation fixes the sign of the surface term in
Eq.~\eqref{eq:magnetic-parent-action}. When comparing equilibrium
solutions with varying $\Lambda$, we keep the resulting horizon
contribution as the surface one-form $\mathcal K_B$ and combine it
with the outer and cosmological-constant contributions.
The sum of these surface and bulk contributions yields the extended boundary identity evaluated below on the static solution family.

For a sphere at fixed $r=R$, radial gauge gives $q_{rr}=-\delta f/f^2$, $q^{rr}=-\delta f$ and $q=-\delta f/f$. We find $\mathcal B_N^r=-nN\delta f/r$. Using the normal directed toward increasing radius, we obtain the surface one-form
\begin{equation}
 \mathcal K_R(\delta)
 =-\frac{n\Omega_nR^{n-1}}{16\pi\GC}\delta f(R)
 =\frac{n\Omega_n}{16\pi\GC}\delta\mu
 +\frac{\Omega_nR^{n+1}}{8\pi\GC(n+1)}\delta\Lambda,
 \qquad
 \widehat H=\frac{n\Omega_n\mu}{16\pi\GC}.
 \label{eq:magnetic-cutoff-charge}
\end{equation}
The unit sphere area is denoted by $\Omega_n$. Pure AdS has the same growing contribution and zero $\mu$, so subtraction at the common asymptotic clock leaves $\delta\widehat H$. Assigning zero energy to AdS fixes the integration constant of this charge. When $\Lambda$ varies, we use the corresponding neighbouring AdS reference, as in Section~\ref{sec:iw}; this retains the same subtraction convention in the canonical and Lorentzian calculations.

At the horizon, regular proper distance $\rho$ gives $N=\kappa\rho+\Order(\rho^3)$. For the normal $s_i$ directed from the horizon into the exterior, we obtain $s_i\mathcal B_N^i=\kappa\sigma^{AB}q_{AB}$ and hence $\mathcal K_B(\delta)=\kappa\delta A/(8\pi\GC)$. This is a surface one-form on the solution family. Its exterior derivative is nonzero when the AdS radius varies independently, as we verify in Section~\ref{sec:schwarzschild}; the integrable energy results from combining the horizon and pressure contributions. The inner outward normal of the exterior slice is $-s_i$, which fixes the sign of the horizon term in the boundary identity.

For the static data, $N\sqrt h=r^n\sqrt\gamma$. We evaluate the reference integral from the AdS origin to $R$ and subtract it from the black-hole integral with lower endpoint $r_h$ and upper endpoint $R$. The resulting finite contribution is $-\Omega_nr_h^{n+1}/(n+1)$. Eq.~\eqref{eq:magnetic-boundary-variation} then yields
\begin{equation}
 \delta\widehat H-\frac{\kappa}{8\pi\GC}\delta A
 =-\frac{\Omega_n r_h^{n+1}}{8\pi\GC(n+1)}\delta\Lambda.
 \label{eq:magnetic-native-identity}
\end{equation}
The horizon coefficient defines $\widehat T=\kappa/(2\pi)$, while the volume coefficient is $\widehat V=\Omega_nr_h^{n+1}/(n+1)$. Together with the hatted entropy and pressure introduced in Section~\ref{sec:contractions}, these coefficients reproduce the conjugate products in Eq.~\eqref{eq:finite-pairings}. Their static boundary identity thus follows from the magnetic action, with the lapse equation and variable-coupling term included throughout. Appendix~\ref{app:threeform} gives a complementary realization in which the cosmological constant is a canonical variable in an enlarged theory.

% 06_schwarzschild.tex
\section{Schwarzschild--AdS thermodynamics and its limiting sectors}
\label{sec:schwarzschild}

\subsection{Thermodynamic potential and independent variations}

We now express the static boundary identity in terms of a thermodynamic potential and its conjugate variables. In four dimensions, the mass charge derived in Section~\ref{sec:magnetic} is $\widehat H=r_0/(2\GC)$, with AdS assigned zero energy. The area entropy follows from Eq.~\eqref{eq:physical-coupling} evaluated at $\GC$, and pressure is defined by Eq.~\eqref{eq:pressure-volume-definition}. Using the horizon area $A=4\pi r_h^2$, we obtain the temperature and enthalpy function
\begin{equation}
 \widehat T=\frac{1+3r_h^2\ell^{-2}}{4\pi r_h},
 \qquad
 \widehat H(\widehat S,\widehat P)
 =\frac12\sqrt{\frac{\widehat S}{\pi\GC}}
      \left(1+\frac83\GC^2\widehat P\widehat S\right).
 \label{eq:sads-enthalpy}
\end{equation}
Here $\widehat S=\pi r_h^2/\GC$ and $\widehat P=3/(8\pi\GC\ell^2)$. Differentiating Eq.~\eqref{eq:sads-enthalpy} at fixed $\widehat P$ gives $\widehat T$. Differentiation at fixed $\widehat S$ gives the positive volume defined in Eq.~\eqref{eq:pressure-volume-definition}, namely the negative of the subtracted integral in Eq.~\eqref{eq:static-volume-subtraction} with $b=1$. The conjugates obtained from the enthalpy derivatives therefore agree with those supplied by the Hamiltonian boundary terms. The pressure definition fixes the subtraction sign in both calculations.

We identify the contributions along the independent directions $(r_h,\ell)$ as
\begin{equation}
 \begin{aligned}
 \widehat T\delta\widehat S
 &=\frac{1}{2\GC}\left(1+\frac{3r_h^2}{\ell^2}\right)\delta r_h,\\
 \widehat V\delta\widehat P
 &=-\frac{r_h^3}{\GC\ell^3}\delta\ell.
 \end{aligned}
 \label{eq:sads-work-components}
\end{equation}
Their sum equals the variation of $r_0/(2\GC)$ through Eq.~\eqref{eq:allowed-variation}. The pressure contribution remains present when the cosmological constant changes, including along variations at fixed horizon area. At a given horizon size, the horizon equation requires the mass coefficient to adjust to a change in the AdS radius. The associated change in energy is therefore accounted for by the pressure term in the extended thermodynamic variation.

Integrability can be checked directly on this two-dimensional family. Differentiating the coefficient of $\delta r_h$ in Eq.~\eqref{eq:sads-work-components} with respect to $\ell$ gives $-3r_h^2/(\GC\ell^3)$, equal to the derivative with respect to $r_h$ of the coefficient of $\delta\ell$. The horizon term by itself has a nonzero exterior derivative when $\ell$ varies. Combining the entropy and pressure terms gives the integrable Hamiltonian differential, which explains the treatment of the horizon contribution as a surface one-form in Section~\ref{sec:magnetic}.

We retain the corresponding scaling identity,
\begin{equation}
 \widehat H=2\widehat T\widehat S-2\widehat P\widehat V,
 \qquad
 H_\lambda=\eta\big(2\widehat T\widehat S-2\widehat P\widehat V\big).
 \label{eq:sads-smarr}
\end{equation}
This Smarr relation follows by dilating $r_h$ and $\ell$ simultaneously at fixed coupling. Each term in the generator-normalized relation carries the common weight $\eta$, in agreement with the differential identity. The agreement of these weights shows that the contraction preserves both the local thermodynamic variation and the integrated relation obtained from a common dilation of the geometric lengths \cite{Kastor:2009wy,Karch_2015,Mancilla:2024spp}.

\subsection{Fixed coupling, the Lorentzian clock and the magnetic representative}

The fixed-coupling sector is obtained with $\alpha=\gamma=0$. In this case, $H_t$, $T_t$ and $V_t$ vanish linearly with $c$, while area entropy and pressure remain fixed. The limiting spatial metric is independent of $t$ and its temporal direction is generated by $\partial_t$, giving a stationary Carroll structure. The Lorentzian norm of this generator tends to zero, as does its Hamiltonian variation at the chosen coupling. Both sides of the limiting thermodynamic identity consequently vanish. The vanishing charge variation is consequently compatible with a surviving stationary symmetry of the limiting geometry, whose temporal generator remains defined.

Ordinary Lorentzian thermodynamics is recovered at $(\alpha,\gamma)=(1,0)$. The generator clock is $s=\tau$, and the temporal metric coefficient is $-f$. With $G=\GC$, the hatted quantities give the usual entropy and volume and satisfy Eq.~\eqref{eq:sads-smarr}. This point is the normalization endpoint of the classification. The Euclidean period transforms inversely to the Hamiltonian, so their thermal weight is invariant under the clock change. Appendix~\ref{app:clock} gives this transformation and relates changes in the clock normalization to the corresponding energy variation.

More generally, at fixed $G$, a positive normalization $\lambda(c)$
retains finite nonzero values of $H_\lambda$, $T_\lambda$ and
$V_\lambda$ when $\lambda(c)c$ approaches a positive finite
constant. For the power law in Eq.~\eqref{eq:scaling-definition},
this constant is unity at $\alpha=1$.
Any other positive finite constant corresponds to a further
constant rescaling governed by Eq.~\eqref{eq:generator-linearity}.

For the finite representative with a fixed generator, we set $(\alpha,\gamma)=(0,1)$. The variables then obey $\eps=c$, $H_t=\widehat H$, $T_t=c\widehat T$, $S=c^{-1}\widehat S$, $V_t=c\widehat V$ and $P=c^{-1}\widehat P$, with the work terms given by Eq.~\eqref{eq:sads-work-components}. The magnetic boundary variation yields the same result because its lapse-weighted spatial measure is $\sqrt f\sqrt h\,\dd^3x=r^2\dd r\,\dd\Omega_2$. Multiplication by $c$ produces the Lorentzian measure associated with $\partial_t$, and the replacement of $G$ by $c\GC$ compensates this factor in the work term.

Through Eq.~\eqref{eq:magnetic-unit-map}, this representative corresponds to the physical family at fixed $G_M$. Its physical energy, pressure and geometric volume have finite limits, while temperature tends to zero and entropy grows as established for magnetic Carroll black holes \cite{Ecker:2023uwm,Tadros:2024}. Eq.~\eqref{eq:finite-pairings} expresses the associated finite boundary mechanics. At every point of the finite branch with the same geometric data, these normalized products coincide; their conversion to a fixed physical clock continues to depend on $\alpha$, as specified in the unit map.

\subsection{Response functions and the meaning of incompressibility}

Response functions require a specified path through the thermodynamic family. At fixed $c$ and coupling, constant pressure means constant $\ell$. Introducing $x=r_h/\ell$, we obtain the heat capacity expressed in hatted variables:
\begin{equation}
 \widehat C_P
 =\widehat T\left(\frac{\partial\widehat S}{\partial\widehat T}\right)_{\widehat P}
 =2\widehat S\,\frac{1+3x^2}{3x^2-1},
 \qquad C_P=c^{-\gamma}\widehat C_P.
 \label{eq:sads-heat-capacity}
\end{equation}
The sign change at $x=1/\sqrt3$ and the pole at the temperature minimum remain those of the ordinary Schwarzschild--AdS family. The contraction changes the overall entropy normalization while preserving these geometric locations. This provides a direct comparison between our extended variables and the specific-heat behaviour analysed in \cite{Tadros:2024}.

For adiabatic variations, fixing $\widehat S$ fixes $r_h$, and therefore fixes $\widehat V$. We obtain zero adiabatic compressibility for the static spherical family. This property holds already at finite $c$ with the corresponding variables. For isothermal variations, we instead find
\begin{equation}
 \widehat\kappa_T
 =-\frac{1}{\widehat V}
     \left(\frac{\partial\widehat V}{\partial\widehat P}\right)_{\widehat T}
 =\frac{24\pi\GC r_h^2}{3x^2-1}.
 \label{eq:sads-compressibility}
\end{equation}
The adiabatic and isothermal responses consequently describe different restrictions on the same equilibrium family. The vanishing of $V_t$ under generator rescaling records a change of conjugate units, while the physical geometric volume remains finite. On the magnetic representative, Eq.~\eqref{eq:physical-thermodynamic-map} gives the physical isothermal compressibility as $\widehat\kappa_T/(\hbar c_*)$, retaining the ensemble condition used in its definition.

A further check is provided by the thermodynamic potential $\mathcal F_\lambda=H_\lambda-T_\lambda S$, for which $\mathcal F_\lambda=\eta r_h(1-x^2)/(4\GC)$. It is finite on the selected branch and vanishes at $r_h=\ell$, as in standard AdS thermodynamics \cite{Kubiznak:2016qmn}. Together, these response functions and the thermodynamic potential show that the geometric locations of the thermodynamic features are preserved, together with the finite normalized work terms. We interpret their physical values using the ensemble conditions and the conversion between generator-normalized and physical units specified above.

% 07_charged.tex
\section{The charged Lorentzian family}
\label{sec:charged}

To include electric work, we consider the Reissner--Nordstr\"om--AdS family \cite{Chamblin:1999tk,Kubiznak:2012wp,Kubiznak:2016qmn}. The Einstein--Maxwell Lagrangian is $\mathbf L_{\rm EM}=(16\pi G)^{-1}(R-2\Lambda-F_{\mu\nu}F^{\mu\nu})\vol$, which fixes the coupling dependence of the electric charge. In the finite-clock coordinate, the static spherical line element and the electromagnetic potential are specified by
\begin{equation}
 f_{\rm RN}(r)=1-\frac{2m}{r}+\frac{q^2}{r^2}+\frac{r^2}{\ell^2},
 \qquad
 m=\frac{r_h}{2}\left(1+\frac{q^2}{r_h^2}+\frac{r_h^2}{\ell^2}\right),
 \qquad
 A=q\left(\frac1{r_h}-\frac1r\right)\dd\tau.
 \label{eq:rn-family}
\end{equation}
Here $r_h$ is the outer horizon, and we initially work at positive surface gravity. The gauge is regular at the bifurcation surface, where the contraction of $A$ with the horizon generator vanishes. The electrostatic potential is therefore defined by the potential difference between infinity and the horizon. With the orientation specified in Section~\ref{sec:setup}, the electric flux gives $Q=q/G$, consistent with the common coupling in the gravitational and Maxwell terms.

The energy $H_\tau=m/G$ is defined with the corresponding electromagnetic boundary prescription. In the horizon-regular gauge, the surface variation of the time translation alone is $\delta(m/G)-\Phi\delta Q$ at infinity. We define the mass generator with a constant compensating gauge parameter $-\Phi$, fixed on the reference solution, which adds $\Phi\delta Q$ to this surface form. Transforming to $A=-q\dd\tau/r$ moves the potential contribution from infinity to the horizon. The thermodynamic law depends on the potential difference between the two boundaries. With the corresponding generator prescription, its energy and electric work therefore agree in both gauges. Under $\tau=ct$, the potential component satisfies $A_t=cA_\tau$ and acquires the same generator factor as temperature when contracted with $\xi_\lambda$.

The independent hatted functions are
\begin{equation}
 \begin{aligned}
 \widehat H_{\rm RN}&=\frac{r_h}{2\GC}
       \left(1+\frac{q^2}{r_h^2}+\frac{r_h^2}{\ell^2}\right),&
 \widehat Q&=\frac q{\GC},\\
 \widehat T_{\rm RN}&=\frac{1-q^2r_h^{-2}+3r_h^2\ell^{-2}}{4\pi r_h},&
 \widehat\Phi&=\frac q{r_h}.
 \end{aligned}
 \label{eq:rn-thermodynamics}
\end{equation}
Entropy, pressure and volume retain the static spherical expressions of Section~\ref{sec:schwarzschild}. Table~\ref{tab:weights} gives the variables at finite $c$, including $\Phi_\lambda=\eps\widehat\Phi$ and $Q=c^{-\gamma}\widehat Q$. We hold the geometric charge $q$ fixed along the contraction curve while allowing $\delta q$ within each finite-$c$ thermodynamic family. This convention allows electric work to contribute along neighbouring equilibrium states while the contraction curve continues to hold the prescribed geometric charge fixed.

We obtain the three work coefficients from elementary parameter variations, keeping the area, electric flux and AdS radius as independent geometric data. For the gauge in Eq.~\eqref{eq:rn-family},
$F=qr^{-2}\dd r\wedge\dd\tau$ and $\star F=q\,\omega_2$, where
$\omega_2$ is the area form of the unit sphere. The electric flux is
therefore independent of the additive horizon potential. At fixed $\GC$,
the area, flux and AdS-radius variations give
\begin{equation}
 \delta\widehat S=\frac{2\pi r_h}{\GC}\delta r_h,\qquad
 \delta\widehat Q=\frac{\delta q}{\GC},\qquad
 \delta\widehat P=-\frac{3}{4\pi\GC\ell^3}\delta\ell.
 \label{eq:rn-parameter-differentials}
\end{equation}
Multiplying these differentials by $\widehat T_{\rm RN}$,
$\widehat\Phi$ and $\widehat V$, respectively, gives the radial, electric
and AdS-radius contributions to the energy variation. In particular, the
negative pressure coefficient follows from the decrease of pressure as
$\ell$ increases at fixed coupling. At finite $c$, the same variations
obey $\delta Q=c^{-\gamma}\delta\widehat Q$ and the corresponding relations
for entropy and pressure. Their conjugates each supply the generator
factor $\eps$, producing the common weight $\eta$ with the contraction
data held fixed.

Differentiation of the energy gives
\begin{equation}
 \delta\widehat H_{\rm RN}
 =\frac{1-q^2r_h^{-2}+3r_h^2\ell^{-2}}{2\GC}\,\delta r_h
   +\frac{q}{\GC r_h}\,\delta q
   -\frac{r_h^3}{\GC\ell^3}\,\delta\ell.
 \label{eq:rn-energy-variation}
\end{equation}
The coefficients are the entropy, electric and pressure work contributions, respectively. Multiplication by $\eta$ gives the full variation of $H_\lambda$, verifying Eq.~\eqref{eq:stationary-extended-law} with $J=0$. Thus the charged family has the same finite branch as the neutral family under the stated fixed-geometric-charge convention. The corresponding integrated identity is
\begin{equation}
 \widehat H_{\rm RN}
 =2\widehat T_{\rm RN}\widehat S
    +\widehat\Phi\widehat Q-2\widehat P\widehat V.
 \label{eq:rn-smarr}
\end{equation}
Both the integrated relation and its differential carry the common contraction weight. In Section~\ref{sec:numerics}, we verify their coefficients symbolically and by numerical differentiation, allowing the entropy, electric and pressure contributions to be checked within the same family.

Holding geometric charge fixed and holding the unrescaled thermodynamic charge $Q$ finite select different contraction paths. To keep $Q$ finite as $G$ tends to zero, we must scale $q$ with $G$. More generally, we can take $q=c^\zeta q_C$, with $q_C$ fixed along $c$, and set $\delta q=c^\zeta\delta q_C$ in each finite-$c$ variation. The electric work then becomes
\begin{equation}
 \Phi_\lambda\delta Q
 =c^{1-\alpha-\gamma+2\zeta}
      \frac{q_C}{\GC r_h}\delta q_C.
 \label{eq:additional-charge-scaling}
\end{equation}
The choice $\zeta=0$ gives the fixed-geometric family considered above. Positive $\zeta$ suppresses both electric work and the charge contribution to the metric, while $\zeta=\gamma$ keeps $Q$ finite. Negative $\zeta$ changes the limiting geometry at fixed $r_h$ and $\ell$ and therefore requires a separate contraction of the geometric data. The charge parameter retained along the contraction curve therefore determines the weight of electric work relative to the entropy and pressure contributions. We specify that choice before evaluating the limiting thermodynamic identity.

At extremality, $q^2=r_h^2(1+3r_h^2/\ell^2)$ and the temperature coefficient in Eq.~\eqref{eq:rn-thermodynamics} vanishes. Restricting Eq.~\eqref{eq:rn-energy-variation} to tangent vectors that preserve this condition leaves the electric and pressure terms as the energy variation. We derive the boundary identity on nonextremal backgrounds and approach the extremal family through this smooth thermodynamic relation. The common factor $\eta$ still governs a nonzero differential on the full family, while the entropy term vanishes after restriction to the extremal subfamily.

The charged family thus tests the thermodynamic contraction of Einstein--Maxwell theory, with its matter contribution retained in the geometric and thermodynamic relations. Its spatial curvature satisfies $R[h]-2\Lambda=2q^2/r^4$, with the electric sector supplying the source. An intrinsic charged Carroll solution therefore requires a compatible contraction of the matter sector, as implemented in the charged constructions of \cite{Ecker:2023uwm}. Section~\ref{sec:magnetic} establishes the vacuum magnetic boundary correspondence for the neutral family. The charged Lorentzian work terms derived here provide the comparison with matter theories in which the electric sector is contracted together with gravity.

% 08_rotating.tex
\section{Kerr--AdS and the rotating thermodynamic variation}
\label{sec:rotating}

\subsection{Asymptotic time and horizon angular velocity}

To examine angular-momentum work and its dependence on the asymptotic frame, we use the four-dimensional Kerr--AdS family \cite{Hawking:1998kw,Gibbons:2004ai,Cvetic:2010jb}. We use Boyer--Lindquist coordinates $(\tau,r,\theta,\phi)$ and the metric
\begin{equation}
 \begin{aligned}
 \dd s^2={}&-\frac{\Delta_r}{\rho^2}
       \left(\dd\tau-\frac{a\sin^2\theta}{\Xi}\dd\phi\right)^2
       +\frac{\rho^2}{\Delta_r}\dd r^2
       +\frac{\rho^2}{\Delta_\theta}\dd\theta^2\\
 &+\frac{\Delta_\theta\sin^2\theta}{\rho^2}
       \left(a\dd\tau-\frac{r^2+a^2}{\Xi}\dd\phi\right)^2,
 \end{aligned}
 \label{eq:kerr-metric}
\end{equation}
where $a$ is the geometric rotation parameter and
\begin{equation}
 \begin{aligned}
 \Delta_r&=(r^2+a^2)(1+r^2\ell^{-2})-2mr,&
 \Delta_\theta&=1-a^2\ell^{-2}\cos^2\theta,\\
 \rho^2&=r^2+a^2\cos^2\theta,&
 \Xi&=1-a^2\ell^{-2}.
 \end{aligned}
 \label{eq:kerr-functions}
\end{equation}
We restrict to $|a|<\ell$ and take $r_h$ to be the largest root of $\Delta_r$. The horizon equation fixes the mass parameter as $m=(r_h^2+a^2)(1+r_h^2\ell^{-2})/(2r_h)$. On the branch with positive Hawking temperature, $r_h$, $a$ and $\ell$ are independent geometric variables, each of which is allowed to vary in the thermodynamic identity.

Thermodynamic energy is defined by a time translation that is nonrotating at infinity. To relate this generator to the Boyer--Lindquist time translation, we set $\psi=\partial_\phi$ and write
\begin{equation}
 \begin{aligned}
 k&=\partial_\tau-\frac{a}{\ell^2}\psi,\\
 \chi&=\partial_\tau+\Omega_{\rm BL}\psi=k+\Omega_\tau\psi,\\
 \Omega_{\rm BL}&=\frac{a\Xi}{r_h^2+a^2},\qquad
 \Omega_\tau=\Omega_{\rm BL}+\frac{a}{\ell^2}
             =\frac{a(1+r_h^2\ell^{-2})}{r_h^2+a^2}.
 \end{aligned}
 \label{eq:kerr-generators}
\end{equation}
In the Boyer--Lindquist expression for $\chi$, the angular velocity is chosen so that the generator has zero norm at the horizon. The coefficient $\Omega_\tau$ measures angular velocity relative to $k$, which is nonrotating at infinity. Holding $\phi_N=\phi+a\tau/\ell^2$ fixed during a time translation produces this generator. Because the transformation of the asymptotic frame depends on $a$ and $\ell$, varying these parameters also varies the relation between the coordinate and nonrotating time translations. We include this dependence when defining the thermodynamic energy and its variation \cite{Gibbons:2004ai}.

\subsection{Thermodynamic volume and variations of all parameters}

At coupling $\GC$, we use the nonrotating-frame functions
\begin{equation}
 \begin{aligned}
 \widehat H_{\rm K}&=\frac{m}{\GC\Xi^2},\\
 \widehat J&=\frac{am}{\GC\Xi^2},\\
 \widehat S_{\rm K}&=\frac{\pi(r_h^2+a^2)}{\GC\Xi},\\
 \widehat T_{\rm K}
 &=\frac{r_h^2-a^2+\ell^{-2}(3r_h^4+a^2r_h^2)}
          {4\pi r_h(r_h^2+a^2)},&&&\\[-1mm]
 \widehat V_{\rm K}
 &=\frac{2\pi(r_h^2+a^2)}{3r_h\Xi^2}
       \left(2r_h^2+a^2-\frac{a^2r_h^2}{\ell^2}\right).&&&
 \end{aligned}
 \label{eq:kerr-thermodynamics}
\end{equation}
Pressure remains $\widehat P=3/(8\pi\GC\ell^2)$, and we set $\widehat\Omega=\Omega_\tau$. Differentiating the thermodynamic functions with respect to the independent geometric parameters gives the full variation. In particular, varying the AdS radius changes $m$ and $\Xi$ as well as pressure, so each dependence must be retained. We verify the coefficients of all parameter variations in $\delta\widehat H_{\rm K}=\widehat T_{\rm K}\delta\widehat S_{\rm K}+\widehat\Omega\delta\widehat J+\widehat V_{\rm K}\delta\widehat P$.

To evaluate the dependence on all three geometric parameters, we differentiate the horizon mass coefficient and $\Xi$ before forming the thermodynamic products. Their variations are
\begin{equation}
 \begin{aligned}
 \delta m={}&\frac12\left(1-\frac{a^2}{r_h^2}
                   +\frac{3r_h^2+a^2}{\ell^2}\right)\delta r_h
   +\frac{a}{r_h}\left(1+\frac{r_h^2}{\ell^2}\right)\delta a
   -\frac{r_h(r_h^2+a^2)}{\ell^3}\delta\ell,\\
 \delta\Xi={}&-\frac{2a}{\ell^2}\delta a
                    +\frac{2a^2}{\ell^3}\delta\ell.
 \end{aligned}
 \label{eq:kerr-geometric-variations}
\end{equation}
The mass, angular momentum and area functions in
Eq.~\eqref{eq:kerr-thermodynamics} consequently have the differentials
\begin{equation}
 \begin{aligned}
 \delta\widehat H_{\rm K}
   &=\frac{\delta m}{\GC\Xi^2}
                    -\frac{2\widehat H_{\rm K}}{\Xi}\delta\Xi,\\
 \delta\widehat J
   &=a\,\delta\widehat H_{\rm K}+\widehat H_{\rm K}\delta a,\\
 \delta\widehat S_{\rm K}
   &=\frac{2\pi}{\GC\Xi}(r_h\delta r_h+a\delta a)
                    -\frac{\widehat S_{\rm K}}{\Xi}\delta\Xi.
 \end{aligned}
 \label{eq:kerr-thermodynamic-differentials}
\end{equation}
Together with the pressure differential in
Eq.~\eqref{eq:rn-parameter-differentials}, these expressions recover the
coefficients of $\delta r_h$, $\delta a$ and $\delta\ell$ in the rotating
identity. The $\delta\Xi$ terms enter the entropy and angular-momentum
variations as well as the energy variation. A change in the AdS radius therefore contributes to several thermodynamic functions through their geometric dependence. Combining these variations retains the single pressure work term conjugate to the energy measured in the nonrotating frame.

To isolate thermodynamic volume, we allow $r_h$ and $a$ to adjust while solving the constraints $\delta\widehat S_{\rm K}=\delta\widehat J=0$ as $\widehat P$ varies. The resulting derivative is expressed by the Jacobian ratio
\begin{equation}
 \widehat V_{\rm K}
 =\frac{\partial(\widehat H_{\rm K},\widehat S_{\rm K},\widehat J)
                     /\partial(r_h,a,\ell)}
        {\partial(\widehat P,\widehat S_{\rm K},\widehat J)
                     /\partial(r_h,a,\ell)}.
 \label{eq:kerr-volume-jacobian}
\end{equation}
Where the denominator is nonzero, this ratio reproduces the volume in Eq.~\eqref{eq:kerr-thermodynamics}. In Section~\ref{sec:numerics}, we test this thermodynamic derivative numerically by solving for neighbouring states at fixed entropy and angular momentum. We then evaluate the energy change per unit pressure change along the resulting constrained family.

The radial volume integral provides a separate geometric quantity with which to compare the thermodynamic derivative. For Eq.~\eqref{eq:kerr-metric}, the metric volume density is $\sqrt{-g}=\rho^2\sin\theta/\Xi$. Using the massless AdS radial primitive in the same coordinates, the negative of the reference-subtracted radial integral gives the contribution
\begin{equation}
 V_{\rm rad}=\frac{4\pi r_h(r_h^2+a^2)}{3\Xi},
 \qquad
 \widehat V_{\rm K}=V_{\rm rad}+\frac{4\pi\GC}{3}a\widehat J.
 \label{eq:kerr-volume-decomposition}
\end{equation}
The additional rotation term accompanies the nonrotating energy and the pressure derivative at fixed entropy and angular momentum. Its relation to the radial volume is part of the extended Kerr thermodynamics analysed in \cite{Cvetic:2010jb}. For the static family, the negative of the subtracted integral in Eq.~\eqref{eq:static-volume-subtraction} gives thermodynamic volume through Eq.~\eqref{eq:pressure-volume-definition}. For Kerr, the chosen frame and energy prescription supply the extra contribution in Eq.~\eqref{eq:kerr-volume-decomposition}. Including this contribution ensures that the volume is conjugate to pressure under the same constraints and energy convention used in the rotating thermodynamic identity.

Frame dependence can also be followed at the level of the thermodynamic differential. Let $\omega_\infty=a/\ell^2$ and define the Boyer--Lindquist energy by $\widehat E_{\rm BL}=\widehat H_{\rm K}-\omega_\infty\widehat J=m/(\GC\Xi)$. Differentiating this relation gives
\begin{equation}
 \delta\widehat E_{\rm BL}
 =\widehat T_{\rm K}\delta\widehat S_{\rm K}
  +\Omega_{\rm BL}\delta\widehat J
  +\widehat V_{\rm K}\delta\widehat P
  -\widehat J\delta\omega_\infty.
 \label{eq:kerr-moving-frame}
\end{equation}
Adding the variation of $\omega_\infty\widehat J$ recovers the law in the nonrotating frame. The final term above therefore accounts for the change in energy induced by varying the angular velocity of the boundary frame. When a fixed physical generator is represented by field-dependent coordinate components, we use Eq.~\eqref{eq:adjusted-surface-form} together with the asymptotic boundary prescription defining $\widehat H_{\rm K}$. For the present family, that prescription is specified by the finite thermodynamic functions in Eq.~\eqref{eq:kerr-thermodynamics}, so the charge variation and thermodynamic derivative use the same energy.

The Kerr Smarr relation is $\widehat H_{\rm K}=2\widehat T_{\rm K}\widehat S_{\rm K}+2\widehat\Omega\widehat J-2\widehat P\widehat V_{\rm K}$. Using the full volume verifies this identity and recovers the static expression as $a\to0$. Entropy, angular momentum and mass carry the inverse coupling weight, while temperature, angular velocity and volume carry the generator weight. Combining these factors gives Eq.~\eqref{eq:work-form-scaling} for the complete rotating differential, including all contributions induced by $\delta\ell$.

\subsection{The contraction and the magnetic momentum sector}

We set $\tau=ct$ while keeping $r_h$, $a$ and $\ell$ fixed along the contraction curve. The metric components obey $g_{tt}=c^2g_{\tau\tau}$, $g_{t\phi}=cg_{\tau\phi}$ and $g_{\phi\phi}=g_{\phi\phi}^{(\tau)}$. We normalize the physical time and horizon generators by $k_\lambda=\eps k$ and $\chi_\lambda=\eps\chi$, respectively. In the $t$ chart these are
\begin{equation}
 \begin{aligned}
 k_\lambda&=c^{-\alpha}\partial_t-\eps\frac{a}{\ell^2}\psi,\\
 \chi_\lambda&=k_\lambda+\Omega_\lambda\psi
             =c^{-\alpha}\partial_t+\eps\Omega_{\rm BL}\psi,
 \qquad \Omega_\lambda=\eps\widehat\Omega.
 \end{aligned}
 \label{eq:kerr-contracted-generators}
\end{equation}
Table~\ref{tab:weights} gives $H_\lambda=\eta\widehat H_{\rm K}$, $J=c^{-\gamma}\widehat J$ and the remaining quantities. On the finite Carroll branch, angular velocity tends to zero while $\Omega_\lambda\delta J$ tends to $\widehat\Omega\delta\widehat J$. The normalized temporal norm carries weight $\eps^2$ and the mixed time-rotation product carries weight $\eps$, wherever their finite-clock coefficients are nonzero. These limits describe the Lorentzian metric and thermodynamic functions along the specified contraction; the canonical momentum determines their relation to the magnetic action.

For that canonical comparison, we write the Lorentzian Kerr metric in ADM form with lapse $N_\tau$ and shift $v_\tau^i$. In the adapted $s$ clock, the lapse is $\eps N_\tau$ and the shift is $\eps v_\tau^i$. Stationarity of the spatial metric then leaves its extrinsic curvature of order unity at fixed $a$. On $G=\eps\GC$, the associated momentum is of order $\eps^{-1}$ wherever the Kerr extrinsic curvature is nonzero. The quadratic momentum contribution to the finite-$\eps$ Hamiltonian therefore remains finite. Thus a fixed-rotation family follows a different momentum scaling from the finite-momentum magnetic sector specified in Section~\ref{sec:magnetic}.

On a Boyer--Lindquist slice, the trace of the extrinsic curvature vanishes, while its squared norm is nonzero for generic rotating data. The Lorentzian scalar constraint consequently gives $R[h]-2\Lambda=K_{ij}K^{ij}$. The static magnetic vacuum constraint requires the curvature value derived in Section~\ref{sec:magnetic}, which identifies the canonical restriction on the limiting geometry. We therefore use Kerr as a rotating Lorentzian thermodynamic test and relate its momentum scaling to the stationary-axisymmetric restrictions of \cite{Kolar:2025rotating}. We interpret the static and topological cases through their respective gravitational solutions and boundary conditions. Finite rotational work remains part of the present thermodynamic comparison, while an intrinsic rotating Carroll interpretation also requires a family satisfying the contracted field equations with its prescribed canonical momentum scaling.

% 09_dimensions.tex
\section{Dependence on spacetime dimension}
\label{sec:dimensions}

We extend the neutral thermodynamic calculation to the spherical Schwarzschild--AdS family in $D\geq4$ dimensions \cite{Kastor:2009wy,Kubiznak:2016qmn}. We write $n=D-2$, and denote the area of the unit $n$-sphere by $\Omega_n=2\pi^{(n+1)/2}/\Gamma((n+1)/2)$. The spatial metric and mass coefficient used in the curvature calculation of Section~\ref{sec:magnetic} have $f=1-\mu r^{1-n}+r^2/\ell^2$, with horizon relation $\mu=r_h^{n-1}(1+r_h^2\ell^{-2})$. We retain $\tau=ct$ and the generator and coupling conventions of Eq.~\eqref{eq:scaling-definition}.

In these dimensions, the gravitational coupling $G$ has units of length to the power $D-2$. Its inverse continues to multiply the dimensionless gravitational action, giving entropy, energy and pressure the same coupling weights as before. The integrated boundary charge derived in Section~\ref{sec:magnetic} supplies the hatted energy with the corresponding sphere-area normalization. Using the same dimensional conventions, we obtain the remaining thermodynamic functions
\begin{equation}
 \begin{aligned}
 \widehat T_D&=\frac{1}{4\pi}
       \left(\frac{D-3}{r_h}+\frac{D-1}{\ell^2}r_h\right),&
 \widehat S_D&=\frac{\Omega_{D-2}r_h^{D-2}}{4\GC},\\
 \widehat P_D&=\frac{(D-1)(D-2)}{16\pi\GC\ell^2},&
 \widehat V_D&=\frac{\Omega_{D-2}r_h^{D-1}}{D-1}.
 \end{aligned}
 \label{eq:dimension-thermodynamics}
\end{equation}
The volume is obtained from the negative of the subtracted integral of $r^{D-2}\dd r\,\dd\Omega_{D-2}$, using the same asymptotic clock and pressure convention as in four dimensions. It acquires a factor $c$ in the original $t$ normalization and a factor $\eps$ for the selected generator. We retain this lapse normalization when comparing the higher-dimensional boundary variation with the thermodynamic derivative.

The radial subtraction follows from the lapse-weighted volume measure in arbitrary dimension. We use the spacetime line element
$\dd s^2=-c^2f\,\dd t^2+f^{-1}\dd r^2+r^2\dd\Omega_{D-2}^2$.
Writing $\omega_{D-2}$ for the unit sphere area form, contraction of its
volume form with the selected time generator gives
$\ii_{\xi_\lambda}\vol=\eps r^{D-2}\dd r\wedge\omega_{D-2}$.
The lapse and radial metric factors cancel in this expression.
Let $\Sigma_R$ run from the black-hole horizon to a cutoff $R$, and let
$\Sigma_{0,R}$ run from the AdS origin to the same cutoff. At the same
asymptotic clock, their integrals are
\begin{equation}
 \begin{aligned}
 \int_{\Sigma_R}\ii_{\xi_\lambda}\vol
    &=\frac{\eps\Omega_{D-2}}{D-1}
                       \left(R^{D-1}-r_h^{D-1}\right),\\
 \int_{\Sigma_{0,R}}\ii_{\xi_\lambda}\vol_0
    &=\frac{\eps\Omega_{D-2}}{D-1}R^{D-1}.
 \end{aligned}
 \label{eq:dimension-cutoff-primitives}
\end{equation}
Subtraction removes the cutoff term. The negative of the finite part
gives the volume in Eq.~\eqref{eq:dimension-thermodynamics} multiplied by
$\eps$; for $\partial_t$, this factor is $c$. The same reference
prescription is applied at each value of $\ell$ when the cosmological
constant varies.

The elementary parameter variations are
\begin{equation}
 \begin{aligned}
 \delta\widehat S_D
    &=\frac{(D-2)\Omega_{D-2}r_h^{D-3}}{4\GC}\delta r_h,\\
 \delta\widehat P_D
    &=-\frac{(D-1)(D-2)}{8\pi\GC\ell^3}\delta\ell.
 \end{aligned}
 \label{eq:dimension-parameter-differentials}
\end{equation}
These expressions identify the independent area and cosmological-constant variations before contraction. Multiplication by their conjugates in Eq.~\eqref{eq:dimension-thermodynamics} gives the two work coefficients, with each geometric direction varied independently at fixed coupling.

We differentiate the energy in the independent variables $(r_h,\ell)$ and obtain
\begin{equation}
 \begin{aligned}
 \delta\widehat H_D
 =\frac{(D-2)\Omega_{D-2}}{16\pi\GC}
 \left\{\left[(D-3)r_h^{D-4}
            +(D-1)\frac{r_h^{D-2}}{\ell^2}\right]\delta r_h
       -\frac{2r_h^{D-1}}{\ell^3}\delta\ell\right\}.
 \end{aligned}
 \label{eq:dimension-energy-variation}
\end{equation}
The coefficient of $\delta r_h$ equals $\widehat T_D\partial_{r_h}\widehat S_D$, and that of $\delta\ell$ equals $\widehat V_D\partial_\ell\widehat P_D$. Their mixed derivatives agree, establishing integrability on the geometric parameter space. The pressure-volume identification therefore has the same form as in four dimensions, with the sphere dimension supplying the appropriate area and volume factors.

The common mixed derivative has value
$-(D-1)(D-2)\Omega_{D-2}r_h^{D-2}/(8\pi\GC\ell^3)$.
It follows both by differentiating the radial coefficient with respect to
$\ell$ and by differentiating the AdS-radius coefficient with respect to
$r_h$. Multiplication by $\eta$, which is constant on this parameter
space, preserves this equality for the complete contracted differential.

We also obtain the dimension-dependent Smarr relation,
\begin{equation}
 (D-3)\widehat H_D
 =(D-2)\widehat T_D\widehat S_D-2\widehat P_D\widehat V_D.
 \label{eq:dimension-smarr}
\end{equation}
To obtain these coefficients, we dilate both geometric lengths at fixed $\GC$. Mass has degree $D-3$, entropy has degree $D-2$, and pressure has degree $-2$. The resulting Eq.~\eqref{eq:dimension-smarr} reduces to the Schwarzschild relation in Eq.~\eqref{eq:sads-smarr} at $D=4$, providing the dimensional comparison with the four-dimensional calculation.

The scalar-curvature calculation in Section~\ref{sec:magnetic} gives $R[h]=2\Lambda$ for $\Lambda=-(D-1)(D-2)/(2\ell^2)$. Mass-dependent terms cancel from this constraint, and the radial and angular lapse equations hold with $N=\sqrt f$. Thus the same static magnetic solution and boundary derivation apply in every dimension considered here. The scalar and lapse equations therefore establish the bulk conditions needed to interpret the dimensional thermodynamic contraction through the magnetic boundary variation.

Applying Table~\ref{tab:weights} to Eq.~\eqref{eq:dimension-thermodynamics} and the energy charge gives the finite-$c$ quantities. Every nonzero work coefficient carries weight $\eta$, so Eq.~\eqref{eq:finite-branch} remains the finite-branch condition independently of $D$. Dimension changes the geometric area and volume factors and the coefficients of the Smarr relation; the common coupling and generator powers determine the contraction. The point $(\alpha,\gamma)=(1,0)$ retains the finite Lorentzian clock, while $(0,1)$ gives the finite-momentum magnetic representative with fixed time generator and coupling $\GC$.

For $D=3$, the energy reference is specified separately. The static metric function is $f=1-\mu+r^2/\ell^2$, with charge $\mu/(8\GC)$ relative to global AdS and conventional BTZ mass $(\mu-1)/(8\GC)$. These energies differ by a constant under the fixed-coupling thermodynamic variation, so they give the same differential identity. In the three-dimensional Smarr relation, the energy coefficient is zero, leaving the entropy and pressure-volume balance prescribed by Eq.~\eqref{eq:dimension-smarr}. The intrinsic Carroll BTZ pressure-volume construction is due to \cite{Kolar:2025rotating}. We use the higher-dimensional family here to develop the common static boundary derivation and the associated normalization analysis.

% 10_numerics.tex
% Numerical records are retained in CQG_revision/verification.
% Plot sources and the additional parameter scans are retained separately.
\section{Numerical verification of the thermodynamic contraction}
\label{sec:numerics}

\subsection{Thermodynamic functions and numerical variations}

We supplement the analytical variations with numerical differentiation of
the thermodynamic functions and plots of their conjugate contributions.
The mass, entropy, pressure, electric charge and angular momentum are
evaluated independently, and their variations are compared using the
temperature and potentials at the reference solution. We retain the
Lorentzian charge and asymptotic-frame conventions of
Refs.~\cite{Gibbons:2004ai,Cvetic:2010jb,Kastor:2009wy}.
This calculation tests the coefficients of the extended thermodynamic
identity, including the variation of the AdS length. The parameter scans
then show how those coefficients contribute to the energy variation as
charge or rotation changes within the chosen family.

We measure lengths in a fixed reference unit $L_0$ and set
$G_C=L_0^{D-2}$. Numerical parameter values below are expressed in these
units, so that each directional work contribution has unit $L_0^{-1}$.
The benchmark solutions have $r_h=1.25$ and $\ell=3$, with $q=0.5$ for
Reissner--Nordstr\"om--AdS and $a=0.6$ for Kerr--AdS. We also evaluate
Schwarzschild--AdS in $D=5,6,8$. All benchmark temperatures are positive.
The charged and rotating inner horizons are approximately $0.167$ and
$0.230$, respectively, while $r_h=1.25$ is the outer horizon in both
families. At the rotating benchmark, $\Xi=0.96$. Every displaced
configuration used in the differentiation retains positive temperature,
and every Kerr--AdS configuration retains $\Xi>0$.

For geometric coordinates $x$ and a tangent $v$, we evaluate the symmetric
difference
\begin{equation}
 \mathcal D_h F(x;v)
 =\frac{F(x+hv)-F(x-hv)}{2h}.
 \label{eq:verification-difference}
\end{equation}
Each coordinate direction is tested with $v_i=x_i e_i$. The combined
tangents are $(0.2r_h,-0.15\ell)$ for the static neutral family,
$(0.2r_h,-0.15\ell,0.25q)$ for the charged family, and
$(0.2r_h,-0.15\ell,0.1a)$ for the rotating family. Their coordinates are
ordered as $(r_h,\ell)$, $(r_h,\ell,q)$ and $(r_h,\ell,a)$, respectively.
We keep $c$, $G_C$ and the contraction exponents fixed within every
difference. Writing the conjugate pairs as
$(I_A,X_A)=(T_\lambda,S),(V_\lambda,P),(\Phi_\lambda,Q),
(\Omega_\lambda,J)$, with the relevant pairs included in each family,
we measure the residual by
\begin{equation}
 \mathcal R_h
 =\frac{\left|\mathcal D_hH_\lambda
       -\sum_A I_A\mathcal D_hX_A\right|}
       {\left|\mathcal D_hH_\lambda\right|
       +\sum_A\left|I_A\mathcal D_hX_A\right|}.
 \label{eq:verification-residual}
\end{equation}
Every work coefficient $I_A$ is evaluated at the reference solution.
The denominator includes the magnitudes of all contributions, so that
cancellation between work terms is included in the error assessment.
We set $\mathcal R_h=0$ when every contribution in its denominator vanishes.

\subsection{Conjugate contributions across charge and rotation}

\begin{figure}[tbp]
\centering
\subfloat[Charged family at fixed reference horizon radius and AdS length.
\label{fig:charged-work-terms}]{\includegraphics[width=0.485\linewidth]{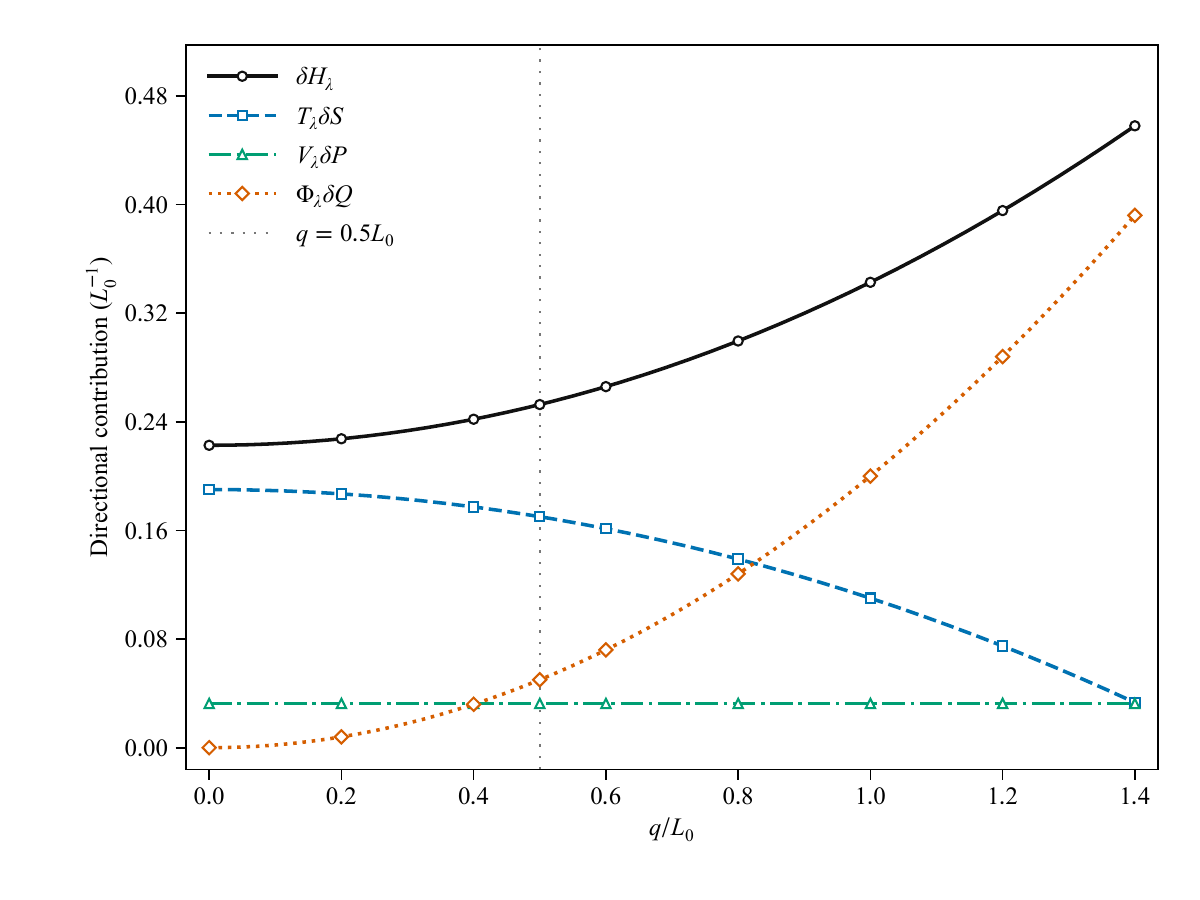}}%
\hfill
\subfloat[Rotating family with energy measured in the nonrotating asymptotic frame.
\label{fig:rotating-work-terms}]{\includegraphics[width=0.485\linewidth]{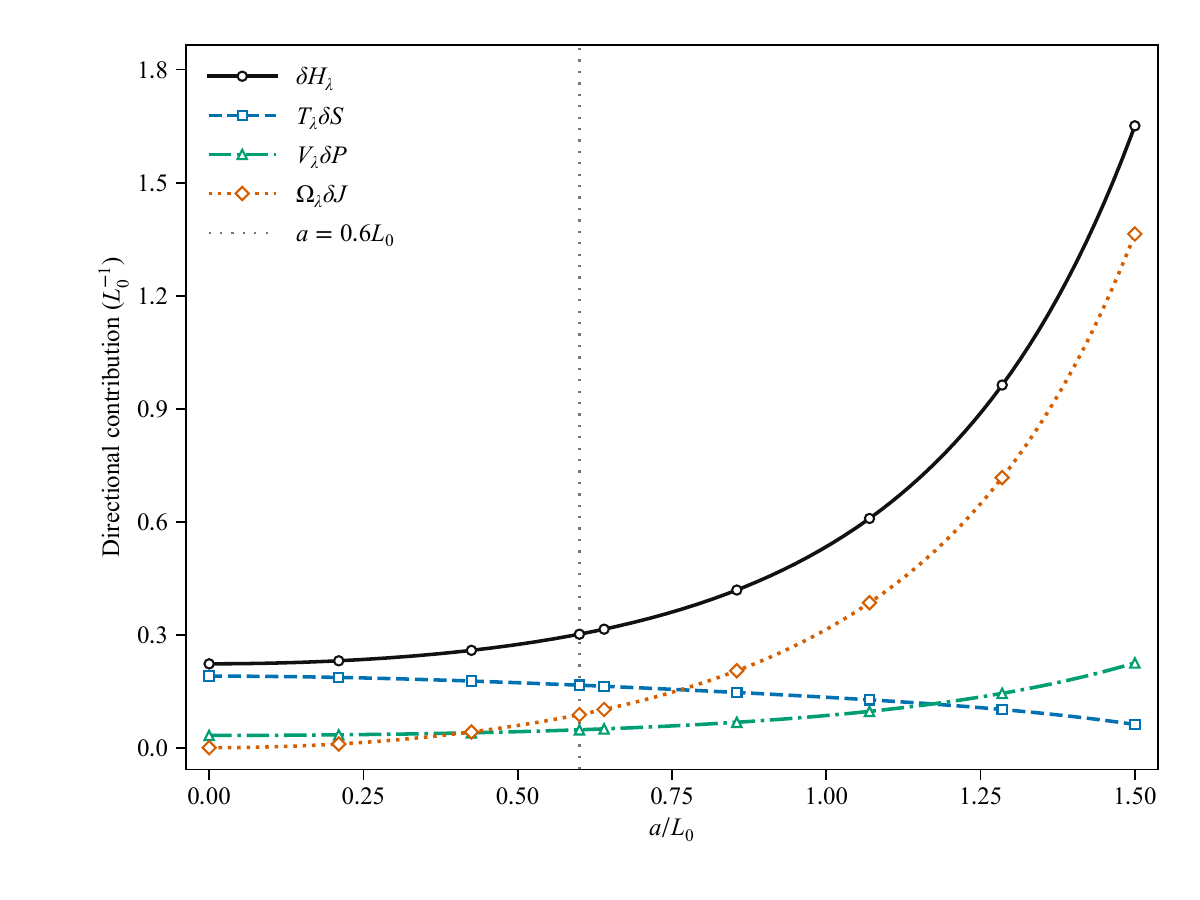}}
\caption{Directional energy variation and its conjugate work contributions
on the finite branch. We use $r_h=1.25$, $\ell=3$ and the combined
tangents defined in the text. The benchmark locations $q=0.5$ and
$a=0.6$ connect these scans to the contraction analysis.
Each curve follows the same prescribed tangent as the energy variation
within its family, so the three work contributions can be compared
directly with their sum.}
\label{fig:work-contributions}
\end{figure}

Figure~\ref{fig:work-contributions} resolves the energy variation into
its entropy, pressure and electric or rotational contributions. We scan
$0\leq q\leq1.4$ in the charged family and $0\leq a\leq1.5$ in the
rotating family, retaining $r_h=1.25$ and $\ell=3$. The scans contain
281 and 301 uniformly spaced reference points, respectively, with
parameter increment $0.005$. Both intervals lie
within the positive-temperature region; along the rotation scan,
$\Xi\geq0.75$. At each reference point we apply the combined tangent
specified above. Thus the horizontal axis changes the reference charge
or rotation, while each plotted variation also includes the prescribed
changes in horizon radius and AdS length. This convention makes the work
contributions comparable throughout each scan. We evaluate the plotted
curves by analytical differentiation and validate the independent
coordinate and combined tangents with symmetric differences at
$h=10^{-4},10^{-5},3\times10^{-6}$. For these scan checks, the
coordinate tangents are $v_i=\max(|x_i|,L_0)e_i$; the combined tangent
retains the prescribed fractional components. These scans provide $6984$
differential checks. At $h=10^{-5}$, their maximum
normalized residuals are below $5\times10^{-10}$ for the charged family
and $3\times10^{-10}$ for the rotating family.

We plot the finite-branch quantities in Eq.~\eqref{eq:finite-pairings}.
For the charged family, Eq.~\eqref{eq:rn-energy-variation} implies that
the electric contribution grows quadratically with $q$, while the
entropy contribution decreases by a term proportional to $q^2$.
The pressure contribution remains constant because the reference horizon
radius, AdS length and fractional AdS-length variation are fixed.
Their sum reproduces the independently differentiated energy. The
benchmark at $q=0.5$ lies in the region where the entropy contribution is
largest and the electric contribution exceeds the pressure contribution.
Along this tangent, the decrease of the entropy contribution is smaller
than the increase of the electric contribution, so the total energy
variation grows across the scan.

Rotation changes all three contributions through the thermodynamic
functions in Eq.~\eqref{eq:kerr-thermodynamics}. In particular, the
pressure term contains the full rotation-dependent volume, and the
angular-momentum term uses the angular velocity relative to the
nonrotating asymptotic frame. At $a=0.6$, the entropy contribution again
dominates, followed by the rotational and pressure contributions.
The rotational contribution becomes dominant as $a$ increases, and
the pressure contribution also exceeds the entropy contribution toward
the upper end of the scan. We evaluate every varied quantity along
the prescribed tangent when comparing the sum with the energy
derivative, including the dependence of $\Xi$ on the AdS length.
The benchmark locations connect these scans to the contraction analysis.

\subsection{Contraction of the work terms and geometric comparison}

\begin{figure}[tbp]
\centering
\includegraphics[width=0.485\linewidth]{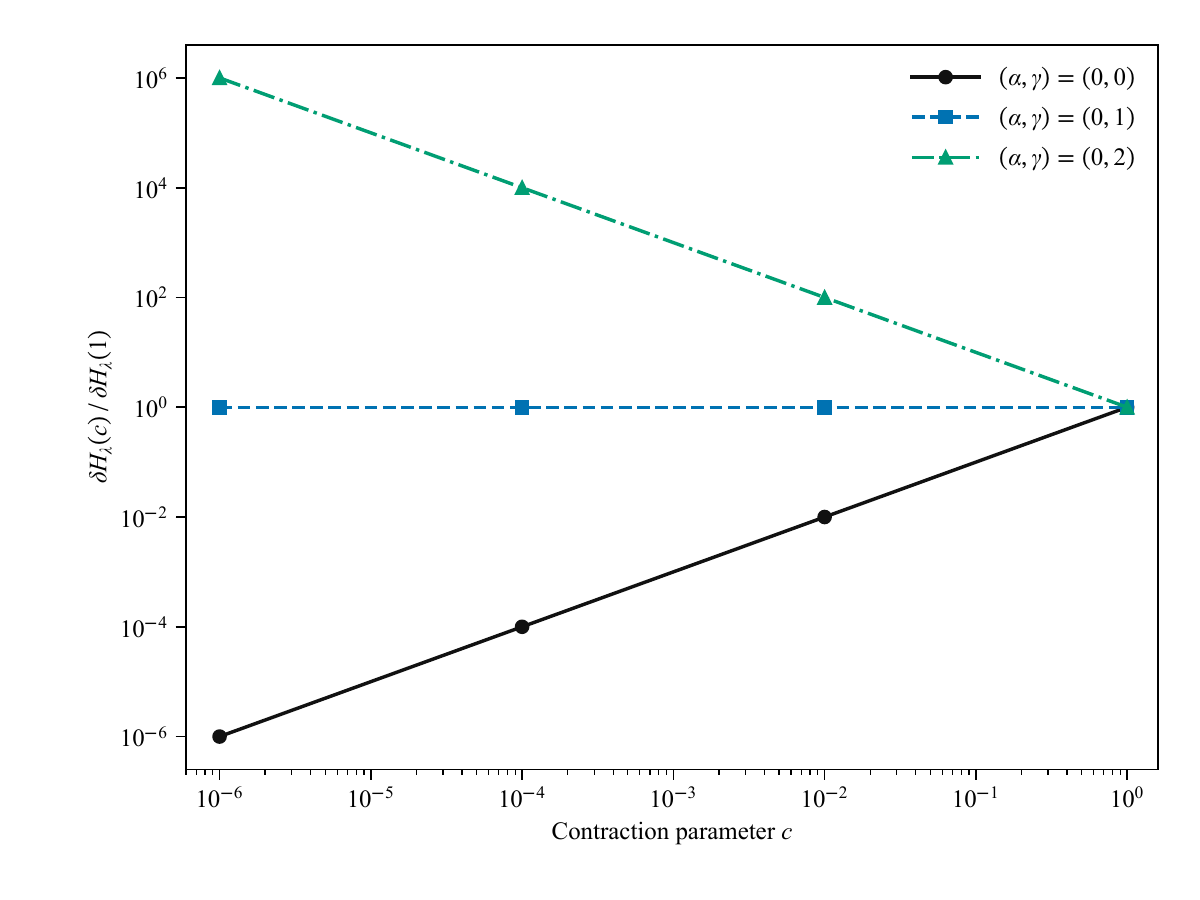}
\caption{Normalized energy variation along the Kerr--AdS contraction.
We use the combined tangent defined in the text and normalize the energy
variation by its value at $c=1$. The curves represent the vanishing,
finite and growing sectors. Each nonzero conjugate work contribution
has the same contraction weight.}
\label{fig:contraction-branches}
\end{figure}

We evaluate $c=1,10^{-2},10^{-4},10^{-6}$ and the exponent pairs
$(\alpha,\gamma)=(0,0),(0,1),(0.5,0.5),(1,0),(0,2)$.
The charged and rotating benchmark contributions remain constant along
the three selected points of $\alpha+\gamma=1$.
Figure~\ref{fig:contraction-branches} shows the
normalized Kerr--AdS energy variation along the combined tangent.
Each nonzero entropy, pressure and rotational work term has the same
ratio to its value at $c=1$, as follows from
Eq.~\eqref{eq:work-form-scaling}. The other families give the same
normalized contraction curves. We therefore display one representative
set while retaining the individual contributions in
Figure~\ref{fig:work-contributions}.

\begin{samepage}
The plotted slopes distinguish the vanishing, finite and growing
thermodynamic differentials. At the smallest sampled $c$, the vanishing
and growing branches differ from their reference values by six orders of
magnitude in opposite directions. The finite branch retains the
reference energy variation. This comparison concerns the common weight
of the thermodynamic one-form; the corresponding gravitational sector
also depends on the canonical momentum condition in
Eq.~\eqref{eq:magnetic-momentum-scaling}.
\par
\end{samepage}

We test the corresponding geometric condition independently at the Kerr
benchmark. At $r=2.5$ and $\theta=\pi/3$, we evaluate the spatial scalar
curvature from Christoffel derivatives and the extrinsic-curvature norm
from the stationary shift. These separate geometric calculations give
$R[h]-2\Lambda$ and $K_{ij}K^{ij}$ at
$\eps=1,10^{-3},10^{-6}$. At each of these three values, both quantities
equal approximately $1.43\times10^{-2}L_0^{-2}$, with an absolute
constraint residual below $5\times10^{-66}L_0^{-2}$ in arithmetic with
65 decimal digits. Their agreement verifies the scalar constraint at the
sampled point. For the neutral comparison, both quantities vanish.
The nonzero Kerr value persists because a common rescaling of the lapse
and shift leaves the extrinsic-curvature norm unchanged. This geometric
test connects the finite Lorentzian rotational work to the momentum
scaling in Section~\ref{sec:rotating}. The static magnetic boundary
calculation instead evaluates the sector with vanishing canonical
momentum, as specified by its field equations.

\subsection{Smarr relation and constrained thermodynamic volume}

We also evaluate the Smarr relation for all six reference families.
We normalize the absolute difference between its two sides by the sum
of the absolute values of the individual terms. The normalized
difference remains below $3\times10^{-16}$ throughout the selected
samples. This agreement checks the algebraic scaling relation across
the sampled dimensions and contraction weights.

We test the rotating volume in Eq.~\eqref{eq:kerr-volume-jacobian}
through a separate constrained calculation. At the displaced AdS
lengths $\ell_\pm=\ell(1\pm h)$, we solve for $r_h$ and $a$ while
keeping $S$ and $J$ at their benchmark values. We then form the ratio
of the corresponding changes in $H_\tau$ and $P$ along that family.
Here $h$ is the fractional change in the AdS length used for the
constrained solve. The calculation uses 65 decimal digits so that the
constraint solve and subtraction errors remain below the finite
difference error. We compare the resulting derivative with the
thermodynamic volume in Eq.~\eqref{eq:kerr-thermodynamics}.
At $h=10^{-5}$, their relative difference is below $7\times10^{-13}$.

The volume is approximately $12L_0^3$, and the relative entropy and
angular-momentum constraint residuals stay below $3\times10^{-60}$.
The volume agreement therefore tests the derivative along the family
of fixed entropy and angular momentum, including the dependence of
$\Xi$ on the AdS length. This constrained calculation checks the
thermodynamic volume independently of its role as the pressure
coefficient in the unconstrained parameter variation.

The supplementary material retains the input parameters, differentiation
scripts and full numerical records, together with the figure sources
and the charge and rotation scans. The three plots resolve the
conjugate contributions and their contraction weights. The curvature, Smarr and constrained-volume
results reported in the text provide additional checks of the geometric
constraint, scaling relation and pressure conjugate, respectively.
Together with the Hamiltonian derivation, these calculations identify
the geometric setting and numerical accuracy of the thermodynamic
contractions considered in this paper.

% 11_conclusion.tex
\section{Conclusion}
\label{sec:conclusion}

In this paper, we have related extended static Carroll boundary mechanics to contractions of Lorentzian AdS black-hole thermodynamics. We connect the physical Newton constant to the coupling in the dimensionless action through a stated unit convention and distinguish thermodynamic variations from motion along the contraction curve. With this convention, the representative defined by a fixed generator reproduces the physical magnetic scaling. Along this path, physical pressure and geometric volume remain finite, temperature vanishes, and entropy diverges at fixed Planck constant. These limits of the individual physical quantities accompany the finite conjugate products obtained from the boundary variation.

The cosmological-constant contribution follows from both the Lorentzian surface identity and the magnetic Hamiltonian variation. A common asymptotic clock and AdS reference fix the bulk and boundary subtractions in the two descriptions. We verify the curvature and lapse equations for the static solution while retaining the canonical momentum sector in the action. The coframe derivation includes the boost connection and reproduces the bulk presymplectic form. In the three-form formulation, a canonical coordinate conjugate to the cosmological constant survives the magnetic contraction. Its evolution and boundary flux together determine the thermodynamic volume associated with the prescribed boundary data.

The energy variation and its conjugate work terms carry the same contraction weight. We use this weight to identify the finite branch and preserve the integrability of the thermodynamic differential at fixed contraction data. The associated hatted variables coincide with the static magnetic boundary quantities; their normalization remains distinct from the limits of the individual unrescaled Lorentzian observables. We also retain the Lorentzian clock endpoint in the classification. This endpoint provides a check of how the thermodynamic identity depends on the chosen time generator and connects the contracted variables to the normalization used in the finite theory.

The charged, rotating and higher-dimensional families provide further tests of the thermodynamic construction. For Kerr--AdS, we include the nonrotating asymptotic frame and the rotational contribution to the thermodynamic volume. The canonical momentum scaling of the family at fixed geometric rotation determines how this Lorentzian test relates to the magnetic theory. Symbolic calculation and numerical differentiation verify the thermodynamic derivatives, including an independent determination of the volume at fixed entropy and angular momentum. Resolving the conjugate work contributions permits comparison of their sum with the energy variation along the prescribed tangent. Together, these checks support the contraction relations on the sampled families.

Further developments include magnetic matter actions with additional charge scalings, boundary conditions for time-dependent AdS perturbations with variable cosmological constant, and higher-derivative contributions to entropy and surface charges. In higher-derivative theories, we would also need to determine how the renormalized bulk volume contracts. This requires following its coupling dependence and reference subtraction together with those of the Wald entropy and surface charges. An independent scaling of the geometric rotation parameter would extend the family considered in Section~\ref{sec:rotating}. We would then compare the angular-momentum work and canonical momentum of that family with the magnetic sector before assigning an intrinsic Carroll interpretation to the resulting limit.

For the finite Carroll branch, we can also seek a boundary Carrollian stress tensor that relates the bulk energy and pressure responses to boundary observables. This would provide a basis for investigating a microscopic interpretation of the finite conjugate products in Carrollian or BMS-invariant field theories. Such a statistical description would need to reproduce the entropy normalization and the conjugate variables identified in the static calculation. The boundary derivation presented in this paper supplies the gravitational quantities required for that comparison and for assessing the thermodynamic interpretations proposed by these extensions within their respective boundary ensembles.

\section*{Data availability statement}
No new data were created or analysed in this study.

\section*{Conflict of interest}
The authors declare no competing interests.

\section*{Ethics statement}
This work did not involve human participants, human data, human tissue, animals, or any activity requiring ethics approval.

\funding{The authors received no specific funding for this work.}

\suppdata{Supplementary material: reproducible thermodynamic and geometric verification, including numerical differentiation, the constrained Kerr volume calculation, and checks of the static magnetic equations.}

\appendix

% 12_appendix_coframe.tex
\section{Coframe action and the canonical symplectic form}
\label{app:coframe}

We use the four-dimensional coframe formulation to identify the connection components that carry canonical momentum. The variables are the Carroll clock form $\tau_C$, spatial coframe $e^a$, rotation connection $\omega^{ab}$ and boost connection $\beta^a$. The coframe description of Carroll geometry and its relation to the magnetic Hamiltonian action were developed in \cite{Hartong:2015xda,Campoleoni:2022wmf}. We define the curvatures by $R^{ab}=\dd\omega^{ab}+\omega^a{}_c\wedge\omega^{cb}$ and $R^a=\dd\beta^a+\omega^a{}_b\wedge\beta^b$ and choose the orientation $\tau_C\wedge e^1\wedge e^2\wedge e^3$. In these conventions, the action form is
\begin{equation}
 \mathbf L_C=\frac{\epsilon_{abc}}{16\pi\GC}
 \left[
 \tau_C\wedge e^a\wedge R^{bc}
 +e^a\wedge e^b\wedge R^c
 -\frac{\Lambda}{3}\tau_C\wedge e^a\wedge e^b\wedge e^c
 \right].
 \label{eq:coframe-full-action}
\end{equation}
The rotational curvature supplies the spatial curvature potential, while the boost curvature supplies the independent momentum sector. Variation with respect to the connections imposes $\dd e^a+\omega^a{}_b\wedge e^b=0$ and $\dd\tau_C+\beta_a\wedge e^a=0$. We solve these equations while retaining the symmetric spatial part of the boost connection as an independent component. Variation of this component imposes zero Carroll extrinsic curvature and thereby relates the coframe constraints to the canonical evolution equation.

We check the normalization by writing the spatial volume form as $e^1\wedge e^2\wedge e^3$. The contraction $\epsilon_{abc}e^a\wedge R^{bc}$ equals $R[h]$ times this volume form, while $\epsilon_{abc}e^a\wedge e^b\wedge e^c$ equals six times the same form. These factors reproduce the coefficient of $R[h]-2\Lambda$ in the canonical action. In the field variation, we retain the boost connection throughout the derivation of the symplectic potential and current, so the normalization check and the canonical momentum sector are treated within the same coframe action.

Varying the clock form while holding $e^a$, $\omega^{ab}$ and
$\beta^a$ independent gives the clock Euler--Lagrange form
\begin{equation}
 \mathbf E_{\tau_C}
 =\frac{\epsilon_{abc}}{16\pi\GC}\,
   e^a\wedge\left(R^{bc}-\frac{\Lambda}{3}e^b\wedge e^c\right).
 \label{eq:coframe-clock-equation}
\end{equation}
On a spatial slice, its pullback is
$(R[h]-2\Lambda)e^1\wedge e^2\wedge e^3/(16\pi\GC)$.
Thus the equation $\mathbf E_{\tau_C}=0$ gives the scalar
constraint in Eq.~\eqref{eq:magnetic-static-equations}.
The independent spatial-coframe and connection variations retain
the boost-curvature term in Eq.~\eqref{eq:coframe-full-action};
after the torsion equations are used, they supply the lapse
equation and canonical momentum sector described in
Section~\ref{sec:magnetic}.

At fixed $\GC$, we integrate the curvature variations by parts and obtain
\begin{equation}
 \begin{aligned}
 \Thgrav_C(\delta)&=
 \frac{\epsilon_{abc}}{16\pi\GC}
 \left[\tau_C\wedge e^a\wedge\delta\omega^{bc}
             +e^a\wedge e^b\wedge\delta\beta^c\right],\\
 \boldsymbol\omega_C(\delta_1,\delta_2)&=
 \frac{\epsilon_{abc}}{16\pi\GC}
 \left[\delta_1(\tau_C\wedge e^a)\wedge\delta_2\omega^{bc}
       +\delta_1(e^a\wedge e^b)\wedge\delta_2\beta^c
       -(1\leftrightarrow2)\right].
 \end{aligned}
 \label{eq:coframe-potential-current}
\end{equation}
We impose time gauge, $\tau_C=N\dd s$, and write the spatial boost connection as $\beta_i{}^a=U^a{}_b e_i{}^b$, with $U_{ab}=U_{ba}$. The temporal component is $\beta_s{}^a=e^{ia}\partial_iN+v^i\beta_i{}^a$. This expression satisfies the clock torsion equation for a spatially varying lapse and therefore includes the lapse of the static solution. With these variables, we identify both the associated momentum and the functional that relates the coframe and canonical potentials:
\begin{equation}
 \pi^{ij}=\frac{\sqrt h}{16\pi\GC}(U^{ij}-h^{ij}U),
 \qquad
 \mathcal F_\Sigma=\frac{1}{16\pi\GC}\int_\Sigma
          \epsilon_{abc}e^a\wedge e^b\wedge\beta^c
  =\frac{1}{8\pi\GC}\int_\Sigma\dd^3x\sqrt h\,U.
 \label{eq:coframe-momentum-functional}
\end{equation}
Here $U^{ij}=e^i{}_ae^j{}_bU^{ab}$ and $U=h_{ij}U^{ij}$. To compare the coframe and canonical potentials, we vary $\mathcal F_\Sigma$ with respect to the coframes and boost connection before substituting the momentum relation. This order of variation retains the contributions needed to relate the two potentials. The identity
\begin{equation}
 -2\epsilon_{abc}\,\delta e^a\wedge e^b\wedge\beta^c
 =\sqrt h(U^{ij}-h^{ij}U)\delta h_{ij}\,\dd^3x,
 \qquad
 \int_\Sigma\Thgrav_C
 =\int_\Sigma\dd^3x\,\pi^{ij}\delta h_{ij}+\delta \mathcal F_\Sigma
 \label{eq:coframe-canonical-reduction}
\end{equation}
then yields Eq.~\eqref{eq:magnetic-canonical-symplectic} when we antisymmetrize a further variation. We impose time gauge on the variations as well as on the fields; the clock contribution to Eq.~\eqref{eq:coframe-potential-current} therefore pulls back to zero on the spatial slice. In both formulations, we accompany any boundary counterterms with the corresponding corner potentials. This prescription keeps the comparison of the bulk potentials consistent with the boundary terms chosen for the variational problem.

We can verify the algebraic identity at a point in an orthonormal frame. Writing $\delta e^a=M^a{}_b e^b$ gives $\delta h_{ab}=M_{ab}+M_{ba}$. The wedge contraction on the left of Eq.~\eqref{eq:coframe-canonical-reduction} equals twice $\Tr(MU)-\Tr(M)\Tr(U)$ times the spatial volume. The contraction of the symmetric metric variation with $U^{ab}-\delta^{ab}U$ gives the same expression. When varying $\mathcal F_\Sigma$, we include the variations of both coframes and of the connection. Their sum gives the field variation $\delta \mathcal F_\Sigma$ that relates the two slice potentials.

In the static gauge, the spatial boost connection vanishes, while its temporal component contributes a spatial divergence to Eq.~\eqref{eq:coframe-full-action}. This divergence is $-2\partial_i(\sqrt h D^iN)/(16\pi\GC)$ and contributes to the boundary terms that enter the comparison of the coframe and canonical actions. The clock torsion equation relates the temporal boost connection to the spatial lapse gradient. Setting the temporal component to zero would therefore also require the lapse gradient to vanish. We retain this component in the static calculation to preserve that torsion equation for the spatially varying lapse.

On the static family, $U^{ij}=\pi^{ij}=0$, and variations tangent to this family satisfy $\delta\pi^{ij}=0$. The pullback of the bulk presymplectic form consequently vanishes on the family. The energy and horizon surface one-forms remain those evaluated in Section~\ref{sec:magnetic}. In the full canonical phase space, we retain momentum variations and obtain the bulk presymplectic form from the action. The boost connection therefore relates the magnetic symplectic structure on the full phase space to its restriction on the static family used in the boundary calculation.

% 13_appendix_threeform.tex
\section{Three-form realization of the cosmological constant}
\label{app:threeform}

In Section~\ref{sec:iw}, we treat $\Lambda$ as a parameter that varies between gravitational theories. By enlarging the field phase space, we can accommodate neighbouring values of $\Lambda$ as solution data within one theory. We use the linear multiplier formulation associated with Henneaux and Teitelboim \cite{Henneaux:1984ji,HenneauxTeitelboim:1989}. Formulations with a quadratic four-form action and membrane transitions provide related mechanisms for a dynamical cosmological constant \cite{Duff:1980qv,Brown:1987dd,Brown:1988kg}. In the linear formulation used here, the multiplier determines the canonical pair that enters the extended boundary variation.

In four dimensions, we introduce a three-form potential $\mathbf A_3$ and write
\begin{equation}
 \mathbf L_{\rm ext}
 =\frac{R-2\Lambda}{16\pi G}\vol
  +\frac{\Lambda}{8\pi G}\dd\mathbf A_3,
 \qquad
 \dd\Lambda=0,
 \qquad \dd\mathbf A_3=\vol.
 \label{eq:threeform-action-equations}
\end{equation}
Variation with respect to $\mathbf A_3$ and $\Lambda$ gives the displayed field equations. The additional term is independent of the metric before the equations of motion are imposed, so metric variation gives the Einstein equation with cosmological constant. The field $\Lambda$ is constant on each solution, while its value can differ between solutions of the enlarged theory. We therefore describe the extended variation within a single field theory, with the cosmological constant treated as solution data.

The top-form variation contributes the potential $\Thgrav_A=\Lambda\delta\mathbf A_3/(8\pi G)$. For a fixed vector $\xi$, the corresponding Noether charge and surface form are
\begin{equation}
 \mathbf Q_{\xi,A}=\frac{\Lambda}{8\pi G}\ii_\xi\mathbf A_3,
 \qquad
 \mathbf k_{\xi,A}=\frac{\delta\Lambda}{8\pi G}\ii_\xi\mathbf A_3,
 \qquad
 \dd\mathbf k_{\xi,A}
 =-\frac{\delta\Lambda}{8\pi G}\ii_\xi\vol.
 \label{eq:threeform-surface-current}
\end{equation}
The last equality follows from Cartan's identity in a stationary gauge, $\Lie_\xi\mathbf A_3=0$. The source term cancels that in Eq.~\eqref{eq:iw-source-identity}, so the complete surface form is closed on solutions of the enlarged theory. When we separate the three-form contribution from this closed form, we recover the pressure work term in the gravitational identity. The coefficient has the same normalization as in the calculation that treats the cosmological constant as an external parameter.

In deriving the surface form, we cancel the terms proportional to $\Lambda\ii_\xi\delta\mathbf A_3$ between $\delta\mathbf Q_{\xi,A}$ and $\ii_\xi\Thgrav_A$. The remaining coefficient multiplies the variation of the integration constant. Both the geometry and the regular potential can vary along the solution family, and this cancellation incorporates their dependent changes into the cosmological-constant work term. We thus retain the volume coefficient as the thermodynamic conjugate to that variation while allowing the potential to remain regular on each member of the family.

We fix the additive potential by requiring regularity at the horizon. For the static solution written in the $\tau$ clock, we denote the area form of the unit sphere by $\omega_2$. With the same asymptotic clock for the black hole and its AdS reference, we choose the potentials as
\begin{equation}
 \mathbf A_3=\frac{r_h^3-r^3}{3}\dd\tau\wedge\omega_2,
 \qquad
 \mathbf A_{3,0}=-\frac{r^3}{3}\dd\tau\wedge\omega_2,
 \qquad
 \int_{S_\infty}^{\ren}\ii_\xi\mathbf A_3
       -\int_B\ii_\xi\mathbf A_3=V_\xi,
 \label{eq:threeform-stationary-potential}
\end{equation}
Here $\xi=b\partial_\tau$, with $b$ held fixed during the variation. The sign of the potential follows from $\dd r\wedge\dd\tau=-\dd\tau\wedge\dd r$. Its coefficient vanishes quadratically in proper distance from the horizon, which ensures regularity at the bifurcation surface. We choose $\mathbf A_{3,0}$ to be regular at the AdS origin and use the common asymptotic clock in the subtraction. The resulting boundary integral is $4\pi b r_h^3/3$. It equals the positive volume defined in Eq.~\eqref{eq:pressure-volume-definition} by the negative of Eq.~\eqref{eq:static-volume-subtraction}. The closed total surface form then yields the gravitational identity with the contribution $-V_\xi\delta\Lambda/(8\pi G)=V_\xi\delta P$.

The canonical pair also determines the boundary ensemble. The potential $\Thgrav_A$ corresponds to fixed three-form boundary data. Adding $-\dd(\Lambda\mathbf A_3)/(8\pi G)$ to the action changes the potential to $-\mathbf A_3\delta\Lambda/(8\pi G)$, which corresponds to fixed $\Lambda$. In comparisons between neighbouring equilibrium solutions, we retain the pressure representation: $\Lambda$ varies, and its volume coefficient enters the work term. We can subsequently impose an ensemble restriction on this identity. The allowed variations and the additive gauge choice in Eq.~\eqref{eq:threeform-stationary-potential} therefore specify separate aspects of the boundary prescription.

On the finite Carroll branch, $\vol=\eps\vol_C$, with $\vol_C=N\sqrt h\,\dd s\wedge\dd^3x$. We set $\mathbf A_3=\eps\mathbf A_C$ and $G=\eps\GC$, which gives finite coefficients for the top-form term and its potential and leaves the equation $\dd\mathbf A_C=\vol_C$. To identify the canonical pair that survives this limit, we decompose the potential as $\mathbf A_C=u\,\dd^3x+\dd s\wedge b_2$, where $b_2=\tfrac12 b^i\epsilon_{ijk}\dd x^j\wedge\dd x^k$. In these variables, the action, symplectic two-form and constraints are
\begin{equation}
 \begin{aligned}
 I_{A,C}&=\frac{1}{8\pi\GC}\int\dd s\,\dd^3x\,
              \Lambda(\partial_su-\partial_ib^i),\\
 \Omega_{A,C}(\delta_1,\delta_2)&=
 \frac{1}{8\pi\GC}\int_\Sigma\dd^3x\,
       (\delta_1\Lambda\,\delta_2u-\delta_2\Lambda\,\delta_1u),\\
 \partial_s\Lambda&=0,\qquad
 \partial_i\Lambda=0,\qquad
 \partial_su-\partial_ib^i=N\sqrt h.
 \end{aligned}
 \label{eq:threeform-carroll-pair}
\end{equation}
We define the subtracted integral of the spatial density $u$ by $\mathcal U=\int_\Sigma^{\ren}u\,\dd^3x$. After imposing the local constraints and identifying configurations related by the allowed three-form gauge transformations, this integral gives the coordinate conjugate to $\Lambda/(8\pi\GC)$. Its increment and the prescribed boundary flux together measure accumulated spacetime volume. In four-dimensional length units, $\mathcal U$ has dimension $L^4$, whereas the thermodynamic volume has dimension $L^3$. The spatial diffeomorphism constraint acquires an additional term proportional to $u\partial_i\Lambda$, which vanishes on the constraint surface. By separating the three-form surface term from the total closed identity, we recover Eq.~\eqref{eq:magnetic-native-identity}. This construction provides a canonical realization of the extended variation used in the external-coupling formulation of the main text.

Integrating the last equation in Eq.~\eqref{eq:threeform-carroll-pair}, we relate the rate $\partial_s\mathcal U$ to the subtracted lapse-weighted spatial volume and the boundary flux of $b^i$. The negative of this spatial volume gives the static thermodynamic coefficient for the selected time generator. In a stationary gauge, the same contribution appears in the boundary potential, as in Eq.~\eqref{eq:threeform-stationary-potential}. To retain the global canonical pair, we fix residual gauge transformations at the boundary. This condition specifies the volume mode being compared together with its allowed boundary data.

% 14_appendix_clock.tex
\section{Changes of clock normalization and thermal weights}
\label{app:clock}

We now allow the normalization parameter to vary between descriptions of the same finite theory. At fixed $c$, let $\xi_\lambda=\lambda\xi_t$ and $H_\lambda=\lambda H_t$. When $\lambda$ varies together with the state, the ordinary differential of the normalized energy gives the relation
\begin{equation}
 \dd H_\lambda=\lambda\dd H_t+H_t\dd\lambda,
 \qquad
 \dd H_\lambda-H_t\dd\lambda
   =T_\lambda\dd S+V_\lambda\dd P
       +\Phi_\lambda\dd Q+\Omega_\lambda\dd J.
 \label{eq:clock-source-variation}
\end{equation}
The coefficient $H_t$ measures the response to a change in the external time normalization: when we rescale the clock, we change the energy measured relative to it. Treating this normalization as an additional thermodynamic state variable requires a boundary ensemble that supplies independent dynamical or source data for the variation. In the main calculation, we prescribe the clock and set $\delta\lambda=0$. The thermodynamic differential therefore acts on the geometric state variables while holding the time normalization fixed.

We implement the same distinction in the surface variation through Eq.~\eqref{eq:adjusted-surface-form}. For a field-dependent rescaling, $\delta\mathbf Q_{\lambda\xi_t}=\lambda\delta\mathbf Q_{\xi_t}+(\delta\lambda)\mathbf Q_{\xi_t}$. Subtracting $\mathbf Q_{\delta\xi}$ removes the generator variation and retains the field variation at the chosen normalization. We thereby recover the identity used in the main text, with the contraction parameter held as prescribed external data. Eq.~\eqref{eq:clock-source-variation} continues to give the ordinary differential relation between energies assigned to different clock normalizations.

For the neutral static solution, we obtain the finite-$c$ Euclidean period from smoothness at the horizon. Under the clock relation in Eq.~\eqref{eq:scaling-definition},
\begin{equation}
 \beta_\tau=c\beta_t,\qquad H_t=cH_\tau,
 \qquad
 \beta_tH_t=\beta_\tau H_\tau.
 \label{eq:thermal-clock-weight}
\end{equation}
The corresponding Boltzmann factors obey
$\exp(-\beta_tH_t)=\exp(-\beta_\tau H_\tau)$.
For the selected generator, we use the adapted clock $s$, for which $\beta_\tau=\eps\beta_s$ and $H_\lambda=\eps H_\tau$, and therefore $\beta_sH_\lambda=\beta_\tau H_\tau$. This equality compares clock normalizations within a single finite-$c$ theory. Along the finite Carroll branch, the coupling also changes, and the unrescaled Euclidean action can grow with the entropy. We use the thermal-weight identity to express invariance under changes of clock within each theory. Convergence of a partition function along the family of theories additionally involves the changing coupling and is therefore a separate question.

In charged and rotating ensembles, the intensive potentials carry the same clock factor. The thermal combination of energy, charge and angular momentum therefore obeys the corresponding identity, provided that we use the boundary frame and gauge prescription defining these quantities. For Kerr, a change of asymptotic frame also contributes the variation in Eq.~\eqref{eq:kerr-moving-frame}. We retain this contribution when comparing the energies assigned in the two frame descriptions.

Throughout the paper, we vary the geometric parameters of a finite-$c$ solution at fixed contraction data, include the cosmological-constant source with the prescribed AdS subtraction, and then take the contraction. For fixed geometric data, this procedure commutes with differentiation of the finite hatted thermodynamic functions. If the horizon size, charge, rotation or AdS scale also changes with $c$, we must specify the additional powers before evaluating the limit. This prescription defines the family of geometries together with the normalization of its thermodynamic variation and states which data remain fixed in that variation.

\bibliographystyle{cqg_references}
\bibliography{reference}
\end{document}